\documentclass{aa} % for a referee version
\usepackage{graphicx}
\usepackage{txfonts}
\usepackage{longtable}
\usepackage{amsmath}
\usepackage{graphicx}
\usepackage{float}
\usepackage{multirow}
\usepackage{subfigure}
\usepackage{rotfloat}
\usepackage[]{hyperref}
\usepackage{breakurl}

\usepackage{array}

\usepackage{booktabs}
\usepackage{lscape}
\usepackage{subfigure}
\usepackage{titlesec}
\usepackage{nameref}
\usepackage{url}

\begin{document}

\title{Standardizing the Gamma-ray burst as a standard candle and applying to the cosmological probes: constraints on the two-component dark energy model}
\author{Jia-Lun Li\inst{1}, Yu-Peng Yang\inst{1}, Shuang-Xi Yi\inst{1}, Jian-Ping Hu\inst{2}, Yan-Kun Qu\inst{1} and Fa-Yin Wang\inst{2,3}}

\institute{$^1$ School of Physics and Physical Engineering, Qufu Normal University, Qufu 273165, China\\
           $^2$ School of Astronomy and Space Science, Nanjing University, Nanjing 210023, China\\
           $^3$ Key Laboratory of Modern Astronomy and Astrophysics (Nanjing University) Ministry of Education,  China\\
\email{yisx2015@qfnu.edu.cn, ypyang@qfnu.edu.cn, fayinwang@nju.edu.cn}}

\markboth{the Gamma-ray burst as a standard candle and applying to the cosmological probes}{}

\abstract{As one of the most energetic and brightest events, gamma-ray bursts (GRBs) have been used as a standard candle for cosmological probe. Based on the relevant features of GRBs light curves, a plateau phase followed a decay phase, we obtain X-ray samples of 31 GRBs and optical samples of 50 GRBs, which are thought to be caused by the same physical mechanism. We standardize GRBs using the two-dimension fundamental plane relation of the rest-frame luminosity of the plateau emission ($L_{b,z}$) and the end time of plateau ($T_{b,z}$) $L_{b,z}-T_{b,z}$, as well as the three-dimension fundamental plane correlation including the peak energy ($E_{p,i}$) $L_{b,z}-T_{b,z}-E_{p,i}$. For the cosmological probes, we consider the $\omega$CDM model in which the dark energy consists of one component, and mainly focus on the $X_1X_2$CDM model in which the dark energy is made up of two independent components. We obtain the constraints on the related parameters of the cosmological models using the type Ia supernovae (SNe Ia) data and selected X-ray and optical samples. For the $X_1X_2$CDM model, we find that the values of the equations of state parameters of two dark energies, $\omega_1$ and $\omega_2$, are very close. We also conduct the comparison between the models using the Bayesian information criterion, and find that the $\omega$CDM model is favoured.}

\authorrunning{Li et al.}
\titlerunning{Standardizing GRBs as a standard candle and applying to the cosmological probes}
\maketitle

\section{Introduction}

~~~~The study of Ia supernovae (SNe Ia) has firstly revealed the accelerating expansion of the universe~(\citealp{1998AJ....116.1009R, 1999ApJ...517..565P}), which is confirmed by several independent cosmological probes, such as the cosmic microwave background (CMB)(\citealp{2003ApJS..148..175S,2014A&A...571A..16P, 2016A&A...594A..13P, 2020A&A...641A...6P} ) and baryon acoustic oscillation (BAO)(\citealp{2005ApJ...633..560E}). It has been considered that an unknown component of the universe contributing significantly to the current cosmic energy budget, named dark energy (DE)~(\citealp{2006IJMPD..15.1753C, 2008ARA&A..46..385F, 2008GReGr..40..301D}), causes the acceleration of the universe. Most cosmological observations agree well with the flat $\Lambda$CDM model. However, other alternative DE models, such as $\omega$CDM, $\phi$CDM, and so on~(\citealp{1988ApJ...325L..17P, 1988PhRvD..37.3406R, 2013PhRvD..88l3513P}), are not ruled out given the precision of current measurements. Many DE models contain only a single component with an equation of state parameter $\omega=p/\rho$. \cite{2007PhRvD..76l3007G} proposed a different DE model~($X_1X_2$CDM model) with two components $\omega_{1}$ and $\omega_{2}$, and they have used the Gamma-ray bursts (GRBs) and SNe Ia data to investigate the proposed DE model. Here we will consider the two components DE model and investigate the constraints on the related cosmological parameters.

Many cosmological probes, such as SNe Ia, CMB, and so on, have been employed for testing cosmological models. The maximum observable redshift of SNe Ia is currently about $z\sim 2.3$~(\citealp{2018ApJ...859..101S}), and CMB gives the meaningful information about $z\sim 1100$. Obviously, there is a redshift gap between SNe Ia and CMB. It has been proposed that GRBs are very promising as a complement to SNe Ia and CMB, providing us with the opportunity to explore the blank history of the universe.

GRBs are the most violent bursts in the universe, with high energy and brightness that occur in a very short period of time~(\citealp{1973ApJ...182L..85K, 2006RPPh...69.2259M, 2009ARA&A..47..567G, 2011A&A...536A..96W, 2015PhR...561....1K}), and are typically splited as long bursts (LGRBs, $T_{90}> 2s$) or short bursts (SGRBs, $T_{90}<2s$) based on the bimodal duration time $T_{90}$~(\citealp{1993ApJ...413L.101K, 2013ApJ...763...15Q}). The maximum observed redshift of GRBs is about $z \sim 9.4$~(\citealp{2011ApJ...736....7C}), and could reach $z \sim 20$ theoretically~(\citealp{2000ApJ...536....1L}). Furthermore, unlike SNe Ia, gamma-ray photons are mostly unaffected by the interstellar medium (ISM) as they travel towards us~(\citealp{2015NewAR..67....1W}). Given these advantages, GRBs become a particularly appealing cosmological probe and can be employed to investigate the characteristics of the universe at high redshift, being expected to have effective constraints on the parameters of cosmological models~(\citealp{2004ApJ...612L.101D, 2004ApJ...613L..13G, 2006MNRAS.369L..37L, 2007ApJ...660...16S, 2008MNRAS.391L...1K, 2011MNRAS.415.3423W, 2013IJMPD..2230028A, 2017IJMPD..2630002W, 2018SSPMA..48c9505W, 2021JCAP...09..042K,2022MNRAS.514.1828D,2022PASJ...74.1095D}).

Compared with SNe Ia, the luminosity of GRBs covers several orders of magnitude and cannot be directly applied to the cosmological probe as a ``standard candle''. Many correlations (e.g., \citealp{2002A&A...390...81A, 2004ApJ...616..331G, 2004ApJ...609..935Y, 2008MNRAS.391L..79D, 2005ApJ...633..611L}) have been proposed for the related parameters of the prompt and afterglow emission of GRBs. Based on these correlations, GRBs can be standardized as a standard candle for the purposes of the cosmological probes~(\citealp{2022MNRAS.516.1386C, 2022MNRAS.512..439C, 2022MNRAS.510.2928C, 2022MNRAS.516.2575J, 2022ApJ...941...84L, 2022ApJ...935....7L,2023MNRAS.521.4406L}). Many attempts have been made to treat GRBs as a standard candle, and then use the correlations to constrain the cosmological parameters.~(\citealp{2009MNRAS.400..775C, 2014ApJ...783..126P, 2019MNRAS.486L..46A, 2019ApJS..245....1T, 2021MNRAS.501.1520C, 2021ApJ...914L..40D, 2021ApJ...920..135X, 2022MNRAS.514.1828D, 2022MNRAS.516.2575J, 2023MNRAS.518.2201D, 2023arXiv230716467X}). In this paper, we will investigate the two-parameter correlation between the luminosity ($L_{b,z}$) and the end time of plateau ($T_{b,z}$) $L_{b,z}-T_{b,z}$, and the three-parameter correlation including the peak energy ($E_{p,i}$) $L_{b,z}-T_{b,z}-E_{p,i}$.

With the increased number of GRBs, a subclass of GRBs with a platform phase has been discovered. According to the Swift observations, some GRBs exhibit an X-ray afterglow plateau phase followed by a normal decay phase~(\citealp{2006ApJ...642..354Z, 2006ApJ...642..389N, 2006ApJ...647.1213O, 2007ApJ...670..565L, 2015ApJ...807...92Y,2016ApJS..224...20Y,2023ApJS..265...56L}). Specifically, if the central engine is powered by the rotational energy of a newborn strongly magnetic neutron star, the injection of the energy could result in the plateau phase~(\citealp{1998A&A...333L..87D, 2001ApJ...552L..35Z, 2011MNRAS.413.2031M}). The release of the energy of central engine leads to the spin down of magnetar. The energy released by various radiation processes will cause the decay index in the decay stage of light curve to differ. For instance, if gravitational wave (GW) emission or magnetic dipole (MD) radiation predominate, the X-ray light curve's decay index will be -1 or -2, respectively~(\citealp{1998A&A...333L..87D, 2001ApJ...552L..35Z}). GRBs chosen from the same physical origin are considered to be suitable for cosmological purposes. In previous work, we have selected a class of GRBs with this characteristic and looked into the potential correlations $L_{b,z}-T_{b,z}-E_{\gamma,iso}$ and $L_{b,z}-T_{b,z}-E_{p,i}$~(\citealp{2018ApJ...863...50S, 2023ApJ...953...58L}), where we have used the selected X-ray sample (31 GRBs from~\cite{2022ApJ...924...97W}) and optical sample (50 GRBs from~\cite{2018ApJ...863...50S}). More detailed information can be found from Sects. \ref{Sample selection} and \ref{section:correlations}.

When employing the model-dependent GRBs data, one will inevitably be plagued with the circularity problem~(\citealp{2006NJPh....8..123G}) for constraining the cosmological parameters, and many works have been done for solving this problem~(\citealp{2008MNRAS.391..577A, 2008ApJ...685..354L, 2002ApJ...573...37J, 2003ApJ...593..622J, 2019MNRAS.486L..46A}). Following previous work, here we will calibrate the correlation combining the Hubble parameter data and the Gaussian process (GP) method. The detailed procedure of this method is in Sect. \ref{section:3.2}. In this paper, the two selected sets of samples (X-ray and optical samples) are used and combined with SNe Ia sample to give limiting results for the free parameters of the two-component DE model and the one-component model and compare them after we obtain the calibrated two and three-parameter correlations. We compute the Akaike information criterion (AIC) (\citealp{1974ITAC...19..716A}) and the Bayesian information criterion (BIC) (\citealp{1978AnSta...6..461S}) and compare the $X_1X_2$CDM model with the $\omega$CDM model to determine whether evidence for a two-component DE model is present.

The structure of this paper is arranged as follows. In following section, we will give more details of cosmological model. The selection criteria of GRBs is summarized in Sect. \ref{Sample selection}. In Sect. \ref{section:correlations}, we introduce GRBs correlation and calibration methods. In Sect. \ref{section:4}, we will use the calibrated GRBs and SNe Ia data to get the limits on the related parameters of two components DE model. Conclusions and discussions are given in Sect. \ref{section:5}.

\section{The basic properties of DE cosmological model}
\label{cosmological model}

~~~~Many astronomical observations show that the universe is currently in a phase of accelerated expansion. Although the $\Lambda$CDM model accords well with the majority of the evidence, several observational discrepancies and theoretical issues have been discovered, indicating that other cosmological models cannot be totally ruled out. Most universe models contain only one component of DE, and here we will also investigate whether there is a evidence for the two-component DE model proposed by \cite{2007PhRvD..76l3007G}. In this section, we will review the main points of DE models, and one can refer to \cite{2007PhRvD..76l3007G} for more details of two-component DM model. For any cosmological model, the expansion rate function can be written in the form of

\begin{equation}\label{eq:expansion rate function}
H^2(z) = H_0^2 \Omega(z,\theta),
\end{equation}
where $\theta$ is free parameters of the cosmological model. For $\omega$CDM model, $\Omega(z,\theta)$ can be expressed as

\begin{equation}\label{eq:flatwCDM}
\Omega(z,\theta) = \Omega_{m}(1+z)^3 +\Omega_{k}(1+z)^2 + \Omega_{\Lambda}(1+z)^{3(1+\omega)},
\end{equation}
where $\Omega_{m}$ is matter energy density parameter, $\Omega_{k}$ is space curvature, $\Omega_{\Lambda}$ is the DE density parameter and $\omega$ is the equation of state (EOS) parameter of DE. The current values of $\Omega_{m}$, $\Omega_{k}$ and $\Omega_{\Lambda}$ are related by $\Omega_{m}+\Omega_{k}+\Omega_{\Lambda} = 1$. For flat $\omega$CDM model, the space curvature $\Omega_{k}=0$ and $\Omega(z,\theta)$ can be
written as

\begin{equation}\label{eq:non-flatwCDM}
\Omega(z,\theta) = \Omega_{m}(1+z)^3 + \Omega_{\Lambda}(1+z)^{3(1+\omega)}.
\end{equation}

In spatially flat model, we constrain $\Omega_{m}$, $\omega$ and $H_0$, whereas in a spatially non-flat model, we restrict $\Omega_{m}$, $\Omega_{\Lambda}$, $\omega$ and $H_0$.

Following the previous work given by \cite{2007PhRvD..76l3007G}, we also consider the two components DE cosmological model $X_1X_2$CDM model, and the corresponding EOS are $\omega_1$ and $\omega_2$, respectively. We also assume that there is no interaction between the two components of DE. For flat universe, $\Omega(z,\theta)$ can be written as

\begin{equation}\label{eq:flatXCDM}
\begin{split}
\Omega(z,\theta) = &\Omega_{m0}(1+z)^3 + \Omega_{x1}(1+z)^{3(1+\omega_1)}\\
&+ (1-\Omega_{m0}-\Omega_{x1})(1+z)^{3(1+\omega_2)}.
\end{split}
\end{equation}

For non-flat universe

\begin{equation}\label{eq:non-flatXCDM}
\begin{split}
\Omega(z,\theta) = &(1-\Omega_{m0}-\Omega_{x1}-\Omega_{x2})(1+z)^2 + \Omega_{m0}(1+z)^3\\
& + \Omega_{x1}(1+z)^{3(1+\omega_1)} + (\Omega_{x2})(1+z)^{3(1+\omega_2)},
\end{split}
\end{equation}
where $\Omega_{x1}$ and $\Omega_{x2}$ are the densities of the two independent components of DE. There are five free parameters:
$\Omega_{m0}$, $\Omega_{x1}$, $\omega_1$, $\omega_2$ and $H_0$ in the flat $X_1X_2$CDM model and six in the flat case: $\Omega_{m0}$, $\Omega_{x1}$, $\Omega_{x2}$, $\omega_1$, $\omega_2$ and $H_0$.

\section{Sample selection of GRBs}
\label{Sample selection}

~~~~As previously stated, if the central energy of GRB is a newborn strongly magnetized neutron star, the magnetar's continual injection of energy can cause a plateau phase in the X-ray light curves~(\citealp{1998A&A...333L..87D, 2001ApJ...552L..35Z}).
It's interesting that Swift data has detected a significant fraction of the GRBs display a plateau phase in the X-ray light curves~(\citealp{2006ApJ...642..354Z, 2006ApJ...642..389N, 2006ApJ...647.1213O, 2007ApJ...670..565L, 2015ApJ...807...92Y, 2016ApJS..224...20Y}). By screening GRBs with the platform phase and examining the potential correlation of physical parameters of the afterglow, \cite{2008MNRAS.391L..79D} were able to determine the relation of Dainotti correlation between the platform luminosity ($L_b$) and the end time of platform ($T_b$). There are many studies of using the Dainotti correlation to study the cosmological parameters~(\citealp{2009MNRAS.400..775C, 2010MNRAS.408.1181C, 2013MNRAS.436...82D, 2014ApJ...783..126P, 2015A&A...582A.115I, 2022ApJ...925...15L}).
In general, rotational energy is assumed to constitute a newly-born magnetized neutron star's energy reservoir, and it is released outward in the form of a combination of MD radiation and GW quadrupole emission, causing the magnetar to spin down~(\citealp{2016MNRAS.458.1660L}).
The different decaying indices of the X-ray light curves can be explained by the rotation energy loss of magnetars dominated with different forms. If MD (GW) radiation dominates the spin down, the light curve of the X-ray afterglow has a plateau phase following the decay index of -2 (-1)(\citealp{1998A&A...333L..87D, 2001ApJ...552L..35Z}).
Taking the two different cases into account, the spin down luminosity with time evolution is
\begin{equation}\label{eq:Luminosity}
L(t)=\left\{
\begin{array}{lcl}
L_0\times \frac{1}{(1+t/t_b)^2} & & {\rm MD~dominated} ,\\
L_0\times \frac{1}{(1+t/t_b)^1} & & {\rm GW~dominated},
\end{array}
\right.
\end{equation}
where $L_0$ is the plateau luminosity, $t_b$ is the end time of the plateau. Both of them can be obtained by fitting the light curve of the X-ray afterglow.

SNe Ia is a class of cosmological probe that is currently well-established for constraining cosmological parameters. It has a nearly uniform luminosity with an absolute magnitude of $M\simeq-19.5$~(\citealp{2001LRR.....4....1C}).
As standard candle, only SNe Ia from the same physical mechanism can be utilized. In the same manner, GRBs with the same physical mechanism could be standardized as standard candle.
\cite{2008MNRAS.391L..79D} presented a tight correlation between the luminosity $L_X$ and the end time of plateau $T_a$ ( Dainotti correlation) by employing the samples with a plateau phase in the X-ray afterglow.
Other afterglow correlations have been proposed and applied to the cosmological parameter constraints through a set of GRB samples with the plateau phase characteristics~(\citealp{ 2019ApJS..245....1T,2021ApJ...920..135X, 2022MNRAS.514.1828D, 2022A&A...661A..71H, 2022MNRAS.510.2928C, 2023ApJ...943..126D}).
 Previously, all GRBs with plateaus have been chosen and used in cosmology. On the other hand, the distinct decay index of the decay stage in afterglow indicates that GRBs may be caused by different physical mechanisms (MD or GW dominated)~(\citealp{2021MNRAS.507..730H, 2022ApJ...924...97W, 2023ApJ...953...58L}).

 \cite{2022ApJ...924...97W} have chosen 31 long GRBs (MD-LGRBs) with X-ray plateau phase and dominated by MD radiation as a cosmological distance indicator after standardization using the Danotti correlation. \cite{2021MNRAS.507..730H} utilized the same procedure to screen out two sets of samples, short GRBs with plateau phase and dominated by MD radiation (MD-SGRBs), and long GRBs dominated by GW emission (GW-LGRBs). In this work, we will employ 31 GRBs with a platform in the X-ray light curves and a decay index of -2, which have been chosen by~\cite{2022ApJ...924...97W}.

Note that a part of the optical light curves also show a platform phase and a decay phase~(\citealp{2012ApJ...758...27L}). The plateau phase in optical curve is thought to be produced by the central engine's continual energy injection, and the shallow decay segment is regarded to be from the same physical process~(\citealp{1998PhRvL..81.4301D, 1998A&A...333L..87D, 2001ApJ...552L..35Z, 2006MNRAS.372L..19F, 2007ApJ...670..565L, 2013MNRAS.430.1061R, 2014ApJ...785...74L, 2022ApJ...924...69Y}). The correlations $L_{b,z}-T_{b,z}-E_{\gamma,iso}$ and $L_{b,z}-T_{b,z}-E_{p,i}$ have been fitted by~\cite{2018ApJ...863...50S} using 50 well-sampled GRBs. \cite{2022MNRAS.514.1828D} also discovered that cosmological quantities can be determined as efficiently as X-ray sample using the 3D optical Dainotti correlation. Therefore, in this work, we will also use optical samples of 50 GRBs selected by~\cite{2018ApJ...863...50S}.

\section{The correlations of X-ray and optical samples}
\label{section:correlations}
\subsection{Fitting the $L_{b,z}-T_{b,z}$ correlation}
\label{section:3.1}

~~~~The luminosity at the end time of the plateau phase, $L_{b,z}$, can be written as~(\citealp{2007ApJ...662.1093W, 2008MNRAS.391L..79D, 2010ApJ...722L.215D, 2011ApJ...730..135D})

\begin{equation}\label{eq:L_bz}
L_{b,z} = \frac{4{\pi}d_L^2F_b}{\left(1+z\right)},
\end{equation}
where $d_L$ is the luminosity distance and $F_b$ is the plateau flux at the break time. For the flat case, the luminosity distance $d_L$ is in the form of

\begin{equation}\label{eq:dl_flat}
d_L(z,\theta) = \frac{c\left(1+z\right)}{H_0}{\int_0^z}\frac{dz'}{E(z',\theta)},
\end{equation}
with
\begin{equation}\label{eq:dl_flat_H(z)}
E(z',\theta) = \sqrt{{\Omega_m}{\left(1+z\right)}^3+{\Omega_{\Lambda}}},
\end{equation}
where $H_0$ is the Hubble constant. The parameters $\Omega_m$ and $\Omega_{\Lambda}$ represent the cosmic matter density and DE density, respectively. For GRB samples, it is difficult to directly derive the model-independent correlation $L_{b,z}-T_{b,z}$ from data due to the lack of low-redshift GRBs ($z < 0.1$)~(\citealp{2022ApJ...924...97W}). Therefore, here we have set ${\Omega_m} = 0.315$ and $H_0 = 67.36 ~\rm km~s^{-1}~Mpc^{-1}$~(\citealp{2020A&A...641A...6P}) for the flat cosmological model to calculate the luminosity distance $d_L$ with Eq. \ref{eq:dl_flat}.

\begin{figure*}[ht!]\
\center
\resizebox{80mm}{!}{\includegraphics[]{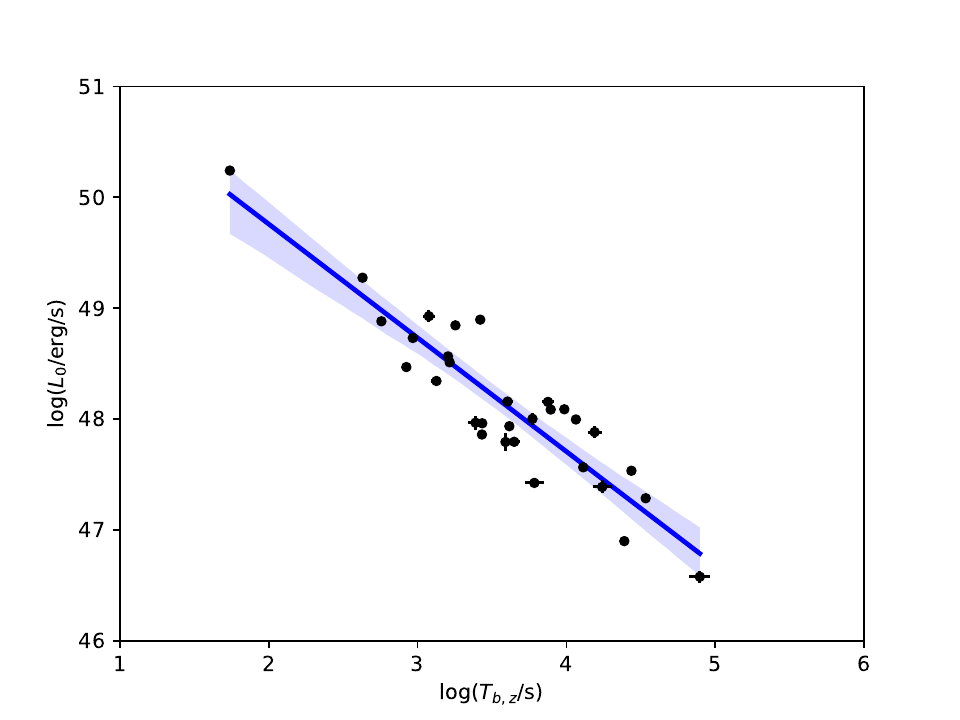}}\resizebox{80mm}{!}{\includegraphics[]{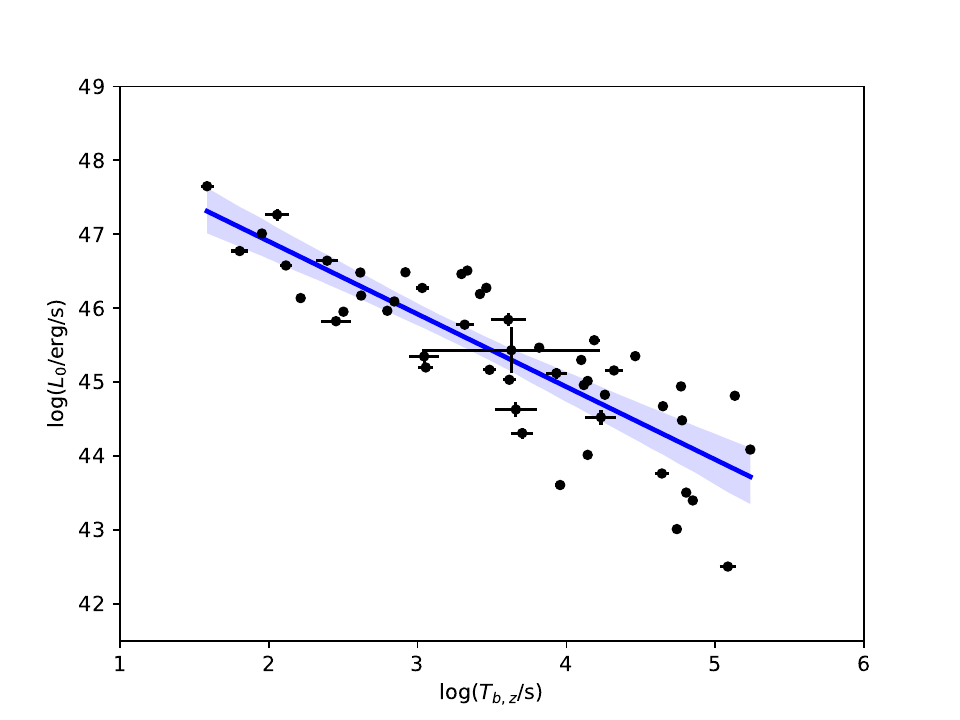}}\\
\resizebox{80mm}{!}{\includegraphics[]{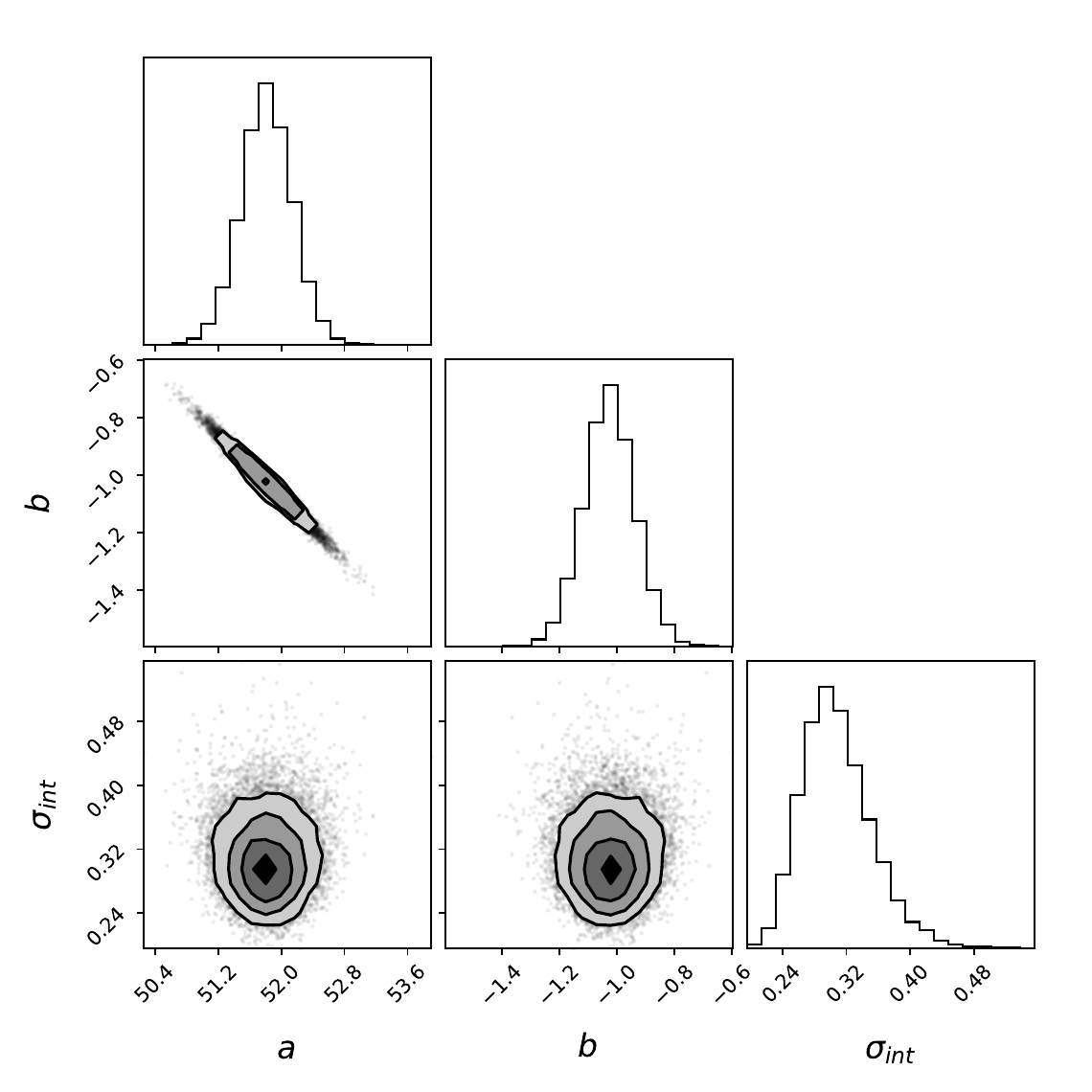}}\resizebox{80mm}{!}{\includegraphics[]{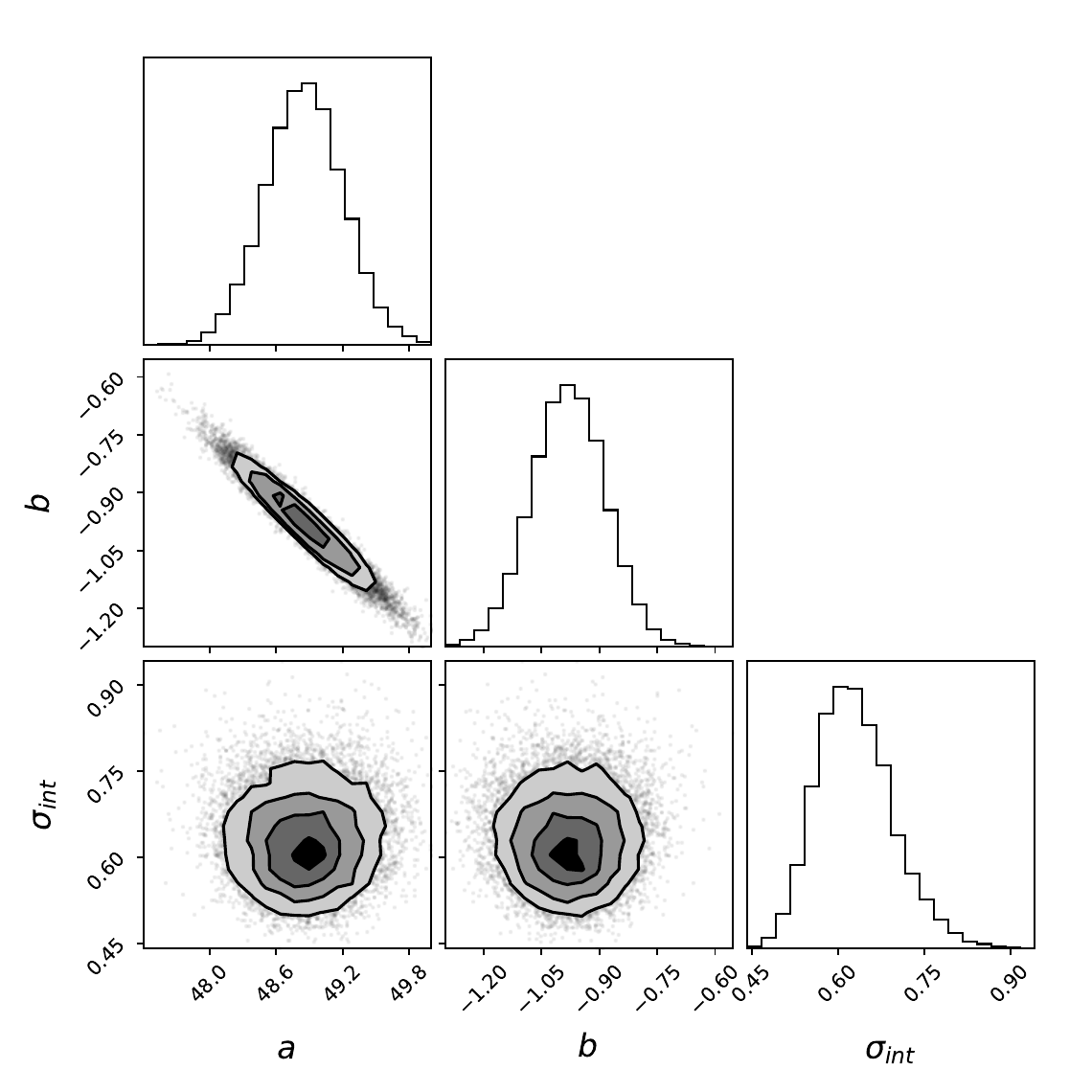}}\\
\caption{The correlation between luminosity $L_{b,z}$ and the end time $T_{b,z}$. Here we have set ${\Omega_m} = 0.315$ and $H_0 = 67.36~{\rm km~s^{-1}~Mpc^{-1}}$ for calculating the luminosity from the measured flux. The data points are the GRBs in X-ray (upper left) and optical samples (upper right). The blue lines correspond to the best fitting values of the data points with a $95\%$ confidence band.  The bottom panels show the 2D posterior contours corner diagrams of the related parameters of the $L_{b,z}-T_{b,z}$ correlation for the X-ray (left panel) and optical sample (right panel).}
\label{fig:L-t(X-ray+optical)}
\end{figure*}

\begin{figure*}[ht!]\
\center
\resizebox{95mm}{!}{\includegraphics[]{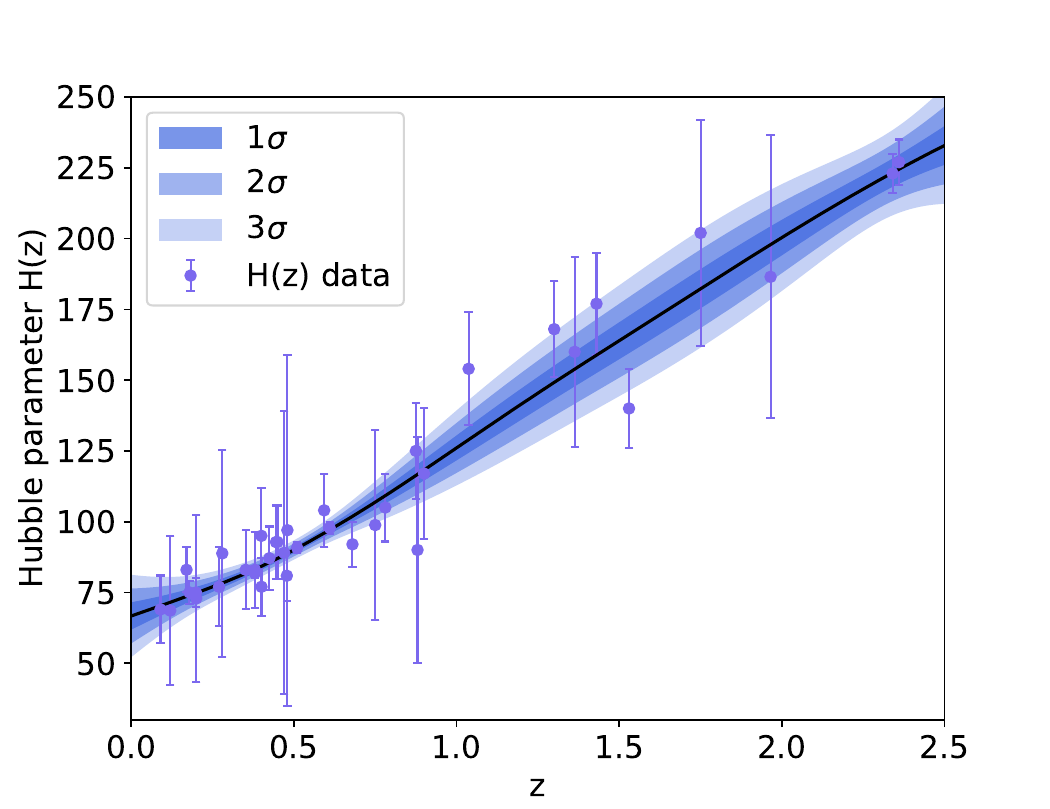}}\\
\caption{The reconstructed smooth $H(z)$ function (the black line) obtained by GP method. The shaded regions correspond to the errors of $1\sigma$, $2\sigma$ and $3\sigma$. The points are 36 $H(z)$ data in the range of $0.07<z< 2.36$. (Source: Fig. 4  in~\cite{2021MNRAS.507..730H}.)}
\label{fig:Hz}
\end{figure*}

The $L_{b,z}-T_{b,z}$ correlation can be written in the following form

\begin{equation}\label{eq:L-T}
{\rm{log}}\left(\frac{L_{b,z}}{\rm{erg~s^{-1}}}\right)= a + b \times {\rm log}\frac{T_{b,z}}{\rm s},
\end{equation}
where $T_{b,z} = t_b / (1+z)$ is the break time of the plateau phase measured in the rest frame. a and b are free parameters. The \texttt{emcee} package~(\citealp{2013PASP..125..306F}) of the Markov Chain Monte Carlo (MCMC) algorithm is used to obtain the best fit values of the parameters a, b and the intrinsic scatter $\sigma_{\rm int}$. The corresponding likelihood function can be written as~(\citealp{2005physics..11182D})

\begin{equation}\label{eq:L-T,likelihood}
\begin{split}
 \mathcal{L}(a,b,\sigma_{\rm int})&\propto\prod_i\frac{1}{\sqrt{\sigma_{\rm int}^2 + \sigma_{y_i}^2 + b^2\sigma_{x_{i}}^2}} \\
 &\times {\rm exp}[-\frac{{(y_i - a - bx_{i})}^2}{2(\sigma_{\rm int}^2 + \sigma_{y_i}^2 + b^2\sigma_{x_{i}}^2)}],
\end{split}
\end{equation}
where $i$ is the serial number of the GRBs for each set of samples. Here we have set $x = {\rm log}(T_{b,z}/{\rm s})$ and $y = {\rm log}(L_{b,z}/{\rm erg~s^{-1}})$. $\sigma_{y_i}$ and $\sigma_{x_{i}}$ are errors of $y_i$ and $x_i$, respectively.

We perform the two-parameter fitting for X-ray (optical) samples and the fitting results are shown in Table \ref{tab:Fitting}. The $L_{b,z}-T_{b,z}$ correlation for the X-ray and optical samples are plotted in Fig. \ref{fig:L-t(X-ray+optical)}.

\subsection{Calibrating the $L_{b,z}-T_{b,z}$ correlation}
\label{section:3.2}
~~~~We fix the cosmological parameters $\Omega_m$ and $H_0$ in deriving the $L_{b,z}-T_{b,z}$ correlation in previous section due to the lack of low-redshift GRBs ($z<0.1$), which is related to the circularity problem~(\citealp{2015NewAR..67....1W}). One needs to derive the correlation independent on the cosmological models in order to use the chosen GRB samples to conduct the cosmological probes. Several approaches have been suggested for resolving the circularity problem~(\citealp{2008A&A...490...31C, 2008MNRAS.391L...1K, 2008ApJ...685..354L, 2011A&A...536A..96W, 2016A&A...585A..68W, 2019MNRAS.486L..46A, 2021ApJ...908..181M, 2022ApJ...924...97W}). Here we follow the model-independent method used by~\cite{2022ApJ...924...97W}, which uses the 36 Hubble parameter data $H(z)$ ($0.07<z<2.36$) compiled by~\cite{2018ApJ...856....3Y}, to calibrate the $L_{b,z}-T_{b,z}$ correlation. Table \ref{tab:H_z} presents the comprehensive 36 $H(z)$ data sets obtained using the following methods:  the cosmic chronometric technique, the BAO signal in galaxy distribution, and BAO signal in Ly$\alpha$ forest distribution alone or cross-correlated with QSOs method. Typically, there are two steps in the calibrating process. First, a continuous function $H(z)$ is constructed from the $H(z)$ data using the GP method, which is then utilized to calibrate the luminosity distances of low redshift GRBs for calculating $L_{b,z}$ according to Eq. \ref{eq:L_bz}. Second, the model-independent distance modulus of GRBs is calculated using the calibrated correlation parameters.

\begin{table}[htbp]
\begin{center}
\caption{Hubble parameter data.}\label{tab:H_z}
\begin{tabular}{|l|c|c|c}
  \hline
  \hline
$z$ &   $H(z) [\rm\,km/s/Mpc]$    & Reference \\ \hline
0.07	&	69	$\pm$	19.6	 & 	(1) \\
0.09	&	69	$\pm$	12	 &	(2) \\
0.12	&	68.6	$\pm$	26.2	 &	(1) \\
0.17	&	83	$\pm$	8	 &	(2) \\
0.179	&	75	$\pm$	4	 &	(3) \\
0.199	&	75	$\pm$	5	 &	(3) \\
0.2	        &	72.9	$\pm$	29.6	 &   (1) \\
0.27	&	77	$\pm$	14	 &   (2) \\
0.28	&	88.8	$\pm$	36.6	 &   (1) \\
0.352	&	83	$\pm$	14	 &   (3) \\
0.38	&	81.9	$\pm$	1.9	 &   (4) \\
0.3802	&	83	$\pm$	13.5	 &   (5) \\
0.4	        &	95	$\pm$	17	 &   (2) \\
0.4004	&	77	$\pm$	10.2	 &   (5) \\
0.4247	&	87.1	$\pm$	11.2	 &   (5) \\
0.4497	&	92.8	$\pm$	12.9	 &   (5) \\
0.47        &       89      $\pm$   50       &   (6) \\
0.4783	&	80.9	$\pm$	9	 &   (5) \\
0.48	&	97	$\pm$	62	 &   (6) \\
0.51	&	90.8	$\pm$	1.9	 &   (4) \\
0.593	&	104	$\pm$	13	 &   (3) \\
0.61	&	97.8	$\pm$	2.1	 &   (4) \\
0.68	&	92	$\pm$	8	 &   (3) \\
0.781	&	105	$\pm$	12	 &   (3) \\
0.875	&	125	$\pm$	17	 &   (3) \\
0.88	&	90	$\pm$	40	 &   (6) \\
0.9	        &	117	$\pm$	23	 &   (2) \\
1.037	&	154	$\pm$	20	 &   (3) \\
1.3	        &	168	$\pm$	17	 &   (2) \\
1.363	&	160	$\pm$	33.6	 &   (8) \\
1.43	&	177	$\pm$	18	 &   (2) \\
1.53	&	140	$\pm$	14	 &   (2) \\
1.75	&	202	$\pm$	40	 &   (2) \\
1.965	&	186.5	$\pm$	50.4	 &   (8) \\
2.34	&	223	$\pm$	7	 &   (9) \\
2.36	&	227	$\pm$	8	 &   (10) \\ \hline
\end{tabular}
\begin{flushleft}
{Reference:(1).\cite{2014RAA....14.1221Z},(2).\cite{2005PhRvD..71l3001S},(3).\cite{2012JCAP...08..006M},(4).\cite{2017MNRAS.470.2617A},
(5).\cite{2016JCAP...05..014M},(6).\cite{2017MNRAS.467.3239R},(7).\cite{2010JCAP...02..008S},(8),\cite{2015MNRAS.450L..16M},
(9).\cite{2015A&A...574A..59D},(10).\cite{2014JCAP...05..027F}
}
\end{flushleft}
\end{center}
\end{table}

We used the public code GaPP ~(\citealp{2006gpml.book.....R, 2012JCAP...06..036S}) to perform the GP regression process and reconstruct the continuous function $H(z)$. Our program employed a \texttt{Matern} kernel and a smoothness parameter of \texttt{nu} =1.5. For the length scale boundaries parameter, we set it to (0.1, 10.0), which ensures that the length scale values remain within this specified interval, resulting in a smoother kernel. We fit GP models with maximum likelihood estimation (MLE) by using the \texttt{fit} function in the package \texttt{GaussianProcessRegressor}\footnote{\url{https://scikit-learn.org/stable/modules/generated/sklearn.gaussian_process.GaussianProcessRegressor.html}}, and the fitted model can be used to make predictions. Then, one can predict the corresponding data using the \texttt{predict} function, and finally get a smooth $H(z)$ reconstruction curve. After the determination of the Hubble parameter $H(z_i)$ using the obtained continuous function $H(z)$, one can obtain the corresponding luminosity distance $d_L(z_i)$ with
 \begin{equation}\label{eq:dL_Hz}
d_L(z) = c(1+z){\int_0^z}\frac{dz'}{H(z')}.
\end{equation}

The reconstructed $H(z)$ is plotted in Fig. \ref{fig:Hz}, from which we can obtain $L_{b,z}$ corresponding to low redshifts for fitting the parameters $a$ and $b$ in the $L_{b,z}-T_{b,z}$ correlation.
In view of the redshift range of used 36 $H(z)$ data, $0.17<z< 2.36$, the reconstructed continuous function can only effectively calibrate GRBs for $z\lesssim 2.50$. Taking this into account, we divided the X-ray and optical samples into two subsamples with a boundary of $z=2.50$, including 14 and 38 GRBs in low-redshift X-ray and optical samples, respectively. The fitting results are shown in Table \ref{tab:Fitting}. It can be found that the calibrated correlation parameters are consistent with that obtained with all data in the X-ray and optical samples, respectively.

\begin{figure*}[ht!]\
\center
\resizebox{80mm}{!}{\includegraphics[]{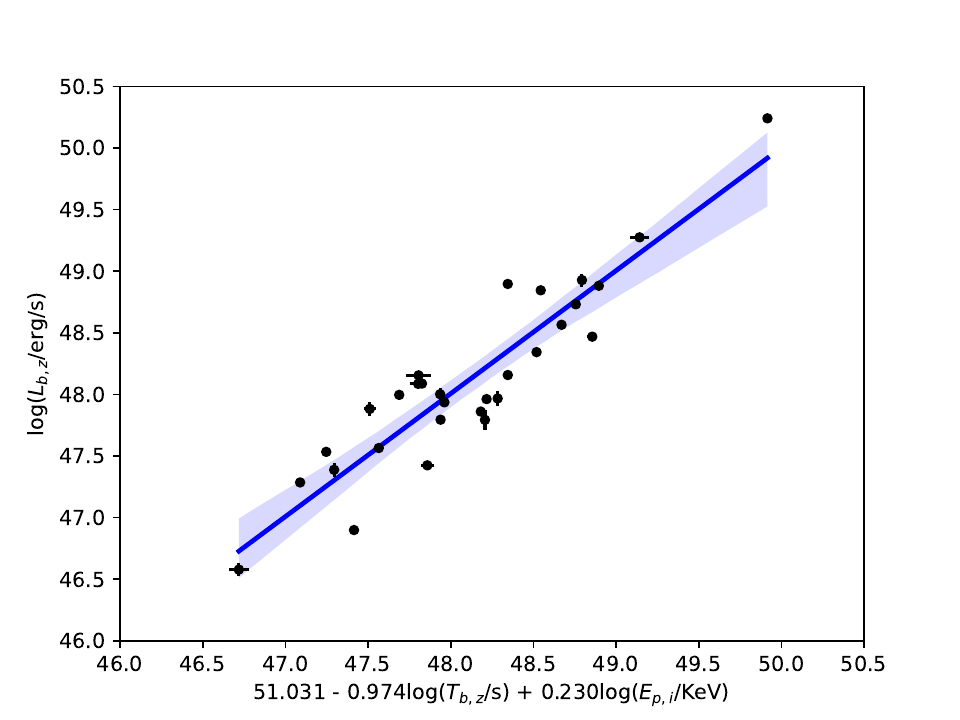}}\resizebox{80mm}{!}{\includegraphics[]{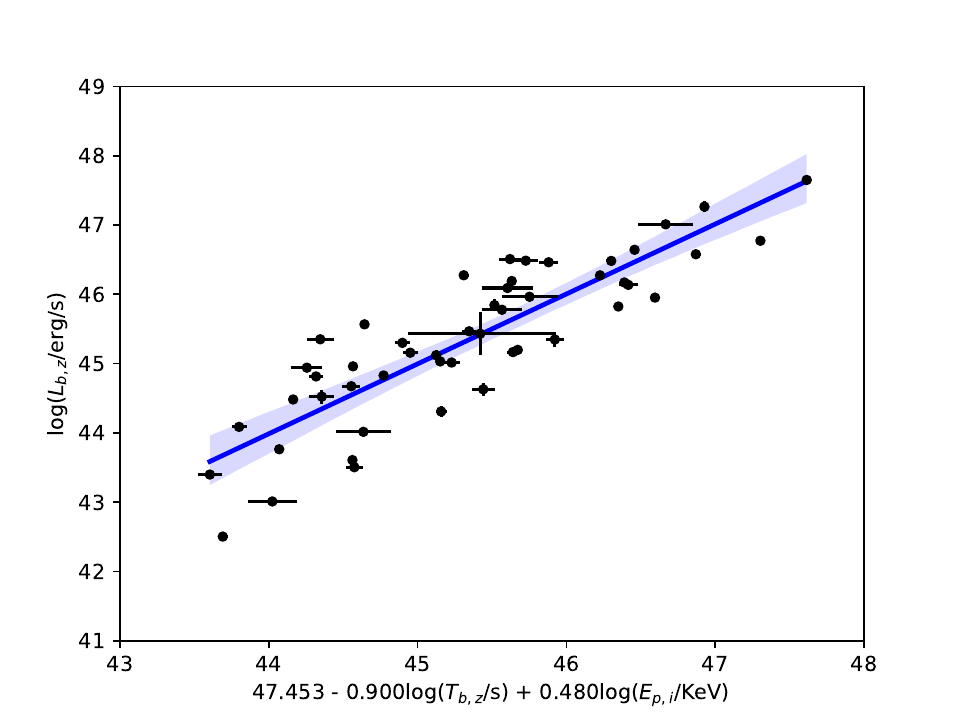}}\\
\resizebox{80mm}{!}{\includegraphics[]{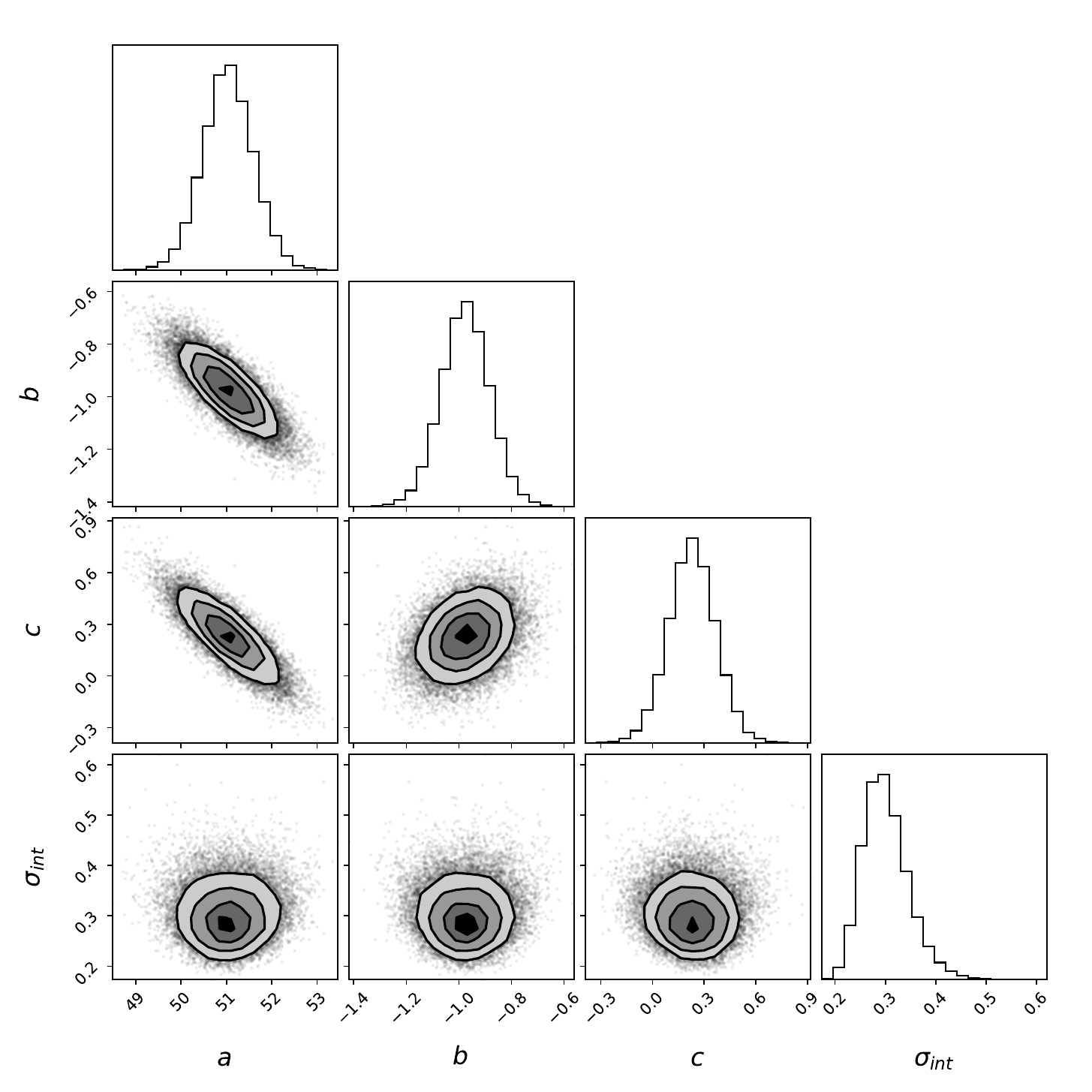}}\resizebox{80mm}{!}{\includegraphics[]{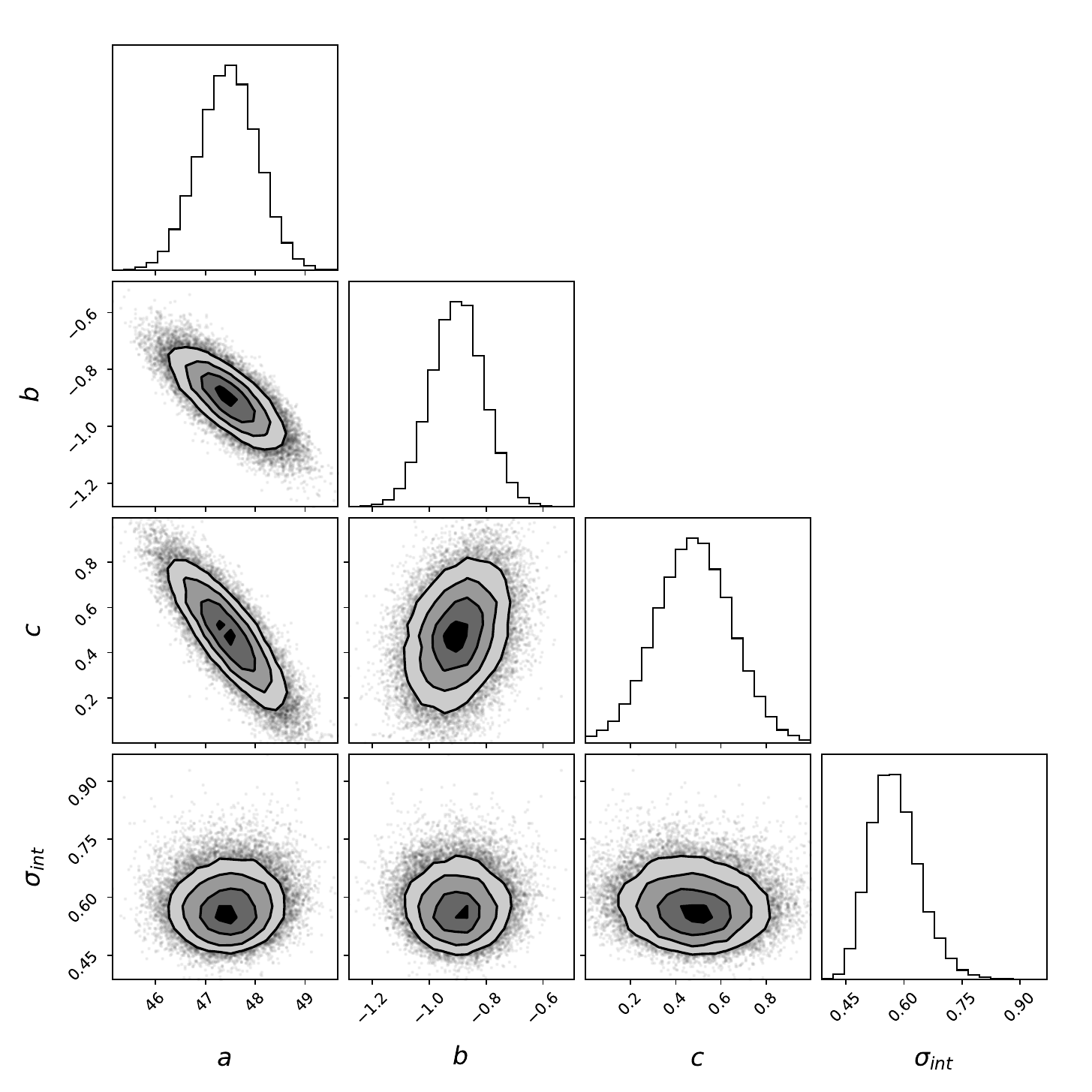}}\\
\caption{The correlation between luminosity $L_{b,z}$, the end time $T_{b,z}$ and the spectral peak energy $E_{p,i}$. Here we have set ${\Omega_m} = 0.315$ and $H_0 = 67.36~{\rm km~s^{-1}~Mpc^{-1}}$ for our calculations. The data points are the GRBs in X-ray (upper left) and optical samples (upper right). The blue line corresponds to the best fitting values of the data points with a $95\%$ confidence band.  The bottom panels show the 2D posterior contours corner diagrams for the X-ray (left panel) and optical samples (right panel).}
\label{fig:L-t-Ep(X-ray+optical)}
\end{figure*}

\subsection{Fitting the $L_{b,z}-T_{b,z}-E_{p,i}$ correlation}
\label{section:3.3}
~~~~Including the spectral peak energy $E_{p,i}$, ~\citealp{2015A&A...582A.115I} proposed the three-parameter correlation $L_{b,z}-T_{b,z}-E_{p,i}$. $E_{p,i}$ is related to the observed peak energy $E_{p,obs}$ in the $\nu F_{\nu}$ spectrum, $E_{p,i} = E_{p,obs} \times (1+z)$. The related data are collected from the literature:~\cite{2018ApJ...863...50S}, \cite{2020MNRAS.492.1919M}, \cite{2021ApJ...920..135X} and \cite{2021MNRAS.508...52L}. \footnote{Among them, GRB 190114A was dropped because of poorly constrained $E_{p,i}$}.  \cite{2018ApJ...863...50S} confirmed the correlation of $L_{b,z}$, $T_{b,z}$, and $E_{p,i}$ using selected optical samples. For the 3D Dinotti correlation, \cite{2022MNRAS.514.1828D} suggested that the optical samples are as effective as X-ray samples for cosmological probes. Based on 121 long GRBs, \cite{2021ApJ...920..135X} pointed out that the de-evolved $L-T-E_p$ correlation can be used as a standard candle to constrain the cosmological parameters.  For our purpose, the three-parameter correlation can be written as

\begin{equation}\label{eq:Ep_relation}
{\rm log}\frac{L_{b,z}}{{\rm erg~s^{-1}}} = a + b{\rm log}\frac{T_{b,z}}{\rm s} + c{\rm log}\frac{E_{p,i}}{\rm keV}.
\end{equation}

The best fit values of the parameters $a$, $b$, $c$ and the internal dispersion $\sigma_{\rm int}$ in the correlation can be obtained from the likelihood function,

\begin{equation}\label{eq:Ep_like}
\begin{split}
 \mathcal{L}(a,b,c,\sigma_{int})&\propto\prod_i\frac{1}{\sqrt{\sigma_{int}^2 + \sigma_{y_i}^2 + b^2\sigma_{x_{1,i}}^2 + c^2\sigma_{x_{2,i}}^2}} \\
 &\times {\rm exp}[-\frac{{y_i - a - bx_{1,i} - cx_{2,i}}^2}{2(\sigma_{int}^2 + \sigma_{y_i}^2 + b^2\sigma_{x_{1,i}}^2 + c^2\sigma_{x_{2,i}}^2)}],
\end{split}
\end{equation}
where $x_1 = {\rm log}(T_{b,z}/s)$, $x_2 = {\rm log}(E_{p,i}/\rm keV)$ and $y = {\rm log}(L_{b,z}/{\rm erg~s^{-1}})$.
The final constraints are shown in Table \ref{tab:Fitting}. The correlation of three parameters and the posterior distribution of the parameters are plotted in  Fig.~\ref{fig:L-t-Ep(X-ray+optical)}. For the same samples, the internal dispersion $\sigma_{\rm int}$ for the three-parameter correlation is smaller than that of the two-parameter correlation, indicating that the $L_{b,z}-T_{b,z} - E_{p,i}$ correlation for optical samples could be applied to the cosmological probes as a standard candle.

\subsection{Calibrating the $L_{b,z}-T_{b,z}-E_{p,i}$ correlation}

~~~~We calibrate the $L_{b,z}-T_{b,z}-E_{p,i}$ correlation using the method described in Sect. \ref{section:3.2} for chosen optical and X-ray samples with low redshift ($z\lesssim 2.50$). The best fitting results are listed in Table \ref{tab:Fitting}. It can be found that the best fit results after calibration are consistent with that obtained by all the GRB X-ray and optical samples. Compared with the case of all GRBs, the internal dispersion of the calibrated two and three-parameter correlations is only marginally increased. In the following section, we will use the calibrated correlations to constrain the related cosmological parameters.

%%%%%%%%%%%%%%%%%%%%%%%%%%%%%%%%%%%%%%%%
\section{Constraints on the cosmological parameters}
\label{section:4}
%%%%%%%%%%%%%%%%%%%%%%%%%%%%%%%%%%%%%%%%%%%%%%%%%%%%%%%%%%%

~~~~The distance modulus for GRBs can be written as

\begin{equation}\label{mu_th}
\mu_{{\rm th}} = 5{\rm log}\frac{d_L}{{\rm Mpc}} + 25 = 5{\rm log}\frac{d_L}{\rm cm} + 97.45.
\end{equation}

After obtaining $d_L$ for the $L_{b,z}-T_{b,z}$ and $L_{b,z}-T_{b,z}-E_{p,i}$ correlations from Eqs. \ref{eq:L_bz}, \ref{eq:L-T}, and \ref{eq:Ep_relation}, we can get the appropriate observed distance modulus for the observed data with above equation. For the two-parameter correlation, the corresponding observed distance modulus is

\begin{equation}\label{eq:mu_bos_L-T}
\mu_{\rm obs} = \frac{5}{2}\Bigg[a + b{\rm log}T_{b,z} - {\rm log}\frac{4{\pi}F_b}{1+z}\Bigg] - 97.45.
\end{equation}

For the errors, following the previous works (e.g., \citealp{2019ApJ...873...39W, 2021ApJ...920..135X, 2022ApJ...924...97W}), we adopt that the physical quantities $L_{b,z}$, $T_{b,z}$ and $E_{p,i}$ are obtained independently from the observational data. Therefore, the uncertainty of the distance modulus can be written as

\begin{equation}\label{eq:mu_obs_err_L-T}
\begin{split}
\sigma_{\rm obs}& = \frac{5}{2}\Bigg[\sigma_{\rm int}^2 + \sigma_a^2 + \sigma_b^2{\rm log}^2_{T_{b,z}} + b^2(\frac{\sigma_{T_{b,z}}}{T_{b,z} {\rm ln}10})^2 + (\frac{\sigma_{F_b}}{F_b {\rm ln}10})^2 \Bigg]^{1/2}.
\end{split}
\end{equation}

For the three-parameter correlation, the observed distance modulus is

\begin{equation}\label{eq:mu_bos_L-T-E}
\mu_{\rm obs} = \frac{5}{2}\Bigg[a + b{\rm log}T_{b,z} + c{\rm log}E_{p,i}-{\rm log}\frac{4{\pi}F_b}{(1+z)}\Bigg] -97.45,
\end{equation}
and the uncertainty of the distance modulus can be written as

\begin{equation}\label{eq:mu_obs_err_L-T-E}
\begin{split}
\sigma_{obs}& = \frac{5}{2}\Bigg[\sigma_{\rm int}^2 + \sigma_a^2 + \sigma_b^2{\rm log}^2_{T_{b,z}} + b^2(\frac{\sigma_{T_{b,z}}}{T_{b,z}{\rm ln}10})^2 + c^2(\frac{\sigma_{E_{p,i}}}{E_{p,i} {\rm ln}10})^2 \\
&+ (\frac{\sigma_{F_b}}{F_b {\rm ln}10})^2 + \sigma_c^2{\rm log}^2_{E_{p,i}}\Bigg]^{1/2}.
\end{split}
\end{equation}

The distance modulus calculated from the best fit of the calibrated correlations can be used to constrain the cosmological models. Specifically, one can constrain
the DE cosmological model by minimizing $\chi^2$,

\begin{equation}\label{eq:chi^2}
\chi^2= \sum_{j=1}^{N}\frac{\left[\mu_{obs}(z) - \mu_{th}(z,\theta)\right]^2}{\sigma_{obs}^2},
\end{equation}
where $N$ is the number of GRBs in each category. The observed distance modulus $\mu_{obs}(z)$ can be calculated from Eqs. \ref{eq:mu_bos_L-T} and \ref{eq:mu_bos_L-T-E}. $\mu_{th}(z,\theta)$ is the theoretical distance modulus, and $\theta$ is the free parameters needed to be constrained.

In general, the luminosity distance can be expressed as

\begin{equation}\label{dl_nonflat}
d_L=\left\{
\begin{array}{lcl}
\frac{c(1+z)}{H_0}(-\Omega_k)^{-\frac{1}{2}}{\rm sin}\left[(-\Omega_k)^{-\frac{1}{2}}{\int_0^z}\frac{dz}{E(z)}\right]
& & {\Omega_k < 0}, \\
\frac{c(1+z)}{H_0}{\int_0^z}\frac{dz}{E(z)}& & {\Omega_k = 0},\\
\frac{c(1+z)}{H_0}\Omega_k^{-\frac{1}{2}}{\rm sinh}\left[\Omega_k^{-\frac{1}{2}}{\int_0^z}\frac{dz}{E(z)}\right] & & {\Omega_k > 0},
\end{array}
\right.
\end{equation}
where $E(z)=H(z)/H_0$ is the expansion rate function of different cosmological models. For $\omega$CDM model, the free parameters are $\Omega_{m0}$, $\Omega_{x0}$, $\omega$ and $H_0$. For $X_1X_2$CDM model, the free parameters are $\Omega_{m0}$, $\Omega_{x1}$, $\Omega_{x2}$, $\omega_1$, $\omega_2$ and $H_0$. We use the \texttt{emcee} package of the MCMC method to find the constraints on those parameters of different models. For the MCMC analysis, we used the following priors for $\omega$CDM model parameters, $\Omega_{m0} \in [0,1]$, $\Omega_{x0} \in [0,1]$, $\omega \in [-5,2]$ and $H_0 \in [65,75]$, and $\Omega_{m0} \in [0,1]$, $\Omega_{x1} \in [0,1]$, $\Omega_{x2} \in [0,1]$, $\omega_1 \in [-5,2]$, $\omega_2 \in [-5,2]$ and $H_0 \in [65,75]$ for $X_1X_2$CDM model. We also included the SNe Ia samples (\citealp{2018ApJ...859..101S}) to get the constraints on the cosmological parameters. It is known that the current limit on $\omega$ is approximately -1, and here we set a slight large range of $\omega_{1,2}$ in order to cover as much parameter space as possible. Many methods can be used to investigate the convergence of MCMC program, and here we use the autocorrelation time for our purposes. In general, a longer autocorrelation time, necessitating a longer MCMC chain, signifies a stronger connection between the samples and sluggish rate of convergence, and the convergent MCMC chains have a short time to explore the whole parameter space. We use the \texttt{sampler.get-autocorr-time} function in \texttt{emcee} to calculate the autocorrelation time. It is found that the autocorrelation time increases with the increase of the number of free parameters. Basically, the chain discarding the preceding values after $\sim 300$ steps starts randomly exploring the entire posterior distribution. For our purposes, we removed the beginning few hundred steps, thinned by about half the autocorrelation time (15 steps), to ensure that the data points cover the full posterior distribution in a short time.

\begin{figure*}[ht!]\
\center
\resizebox{80mm}{!}{\includegraphics[]{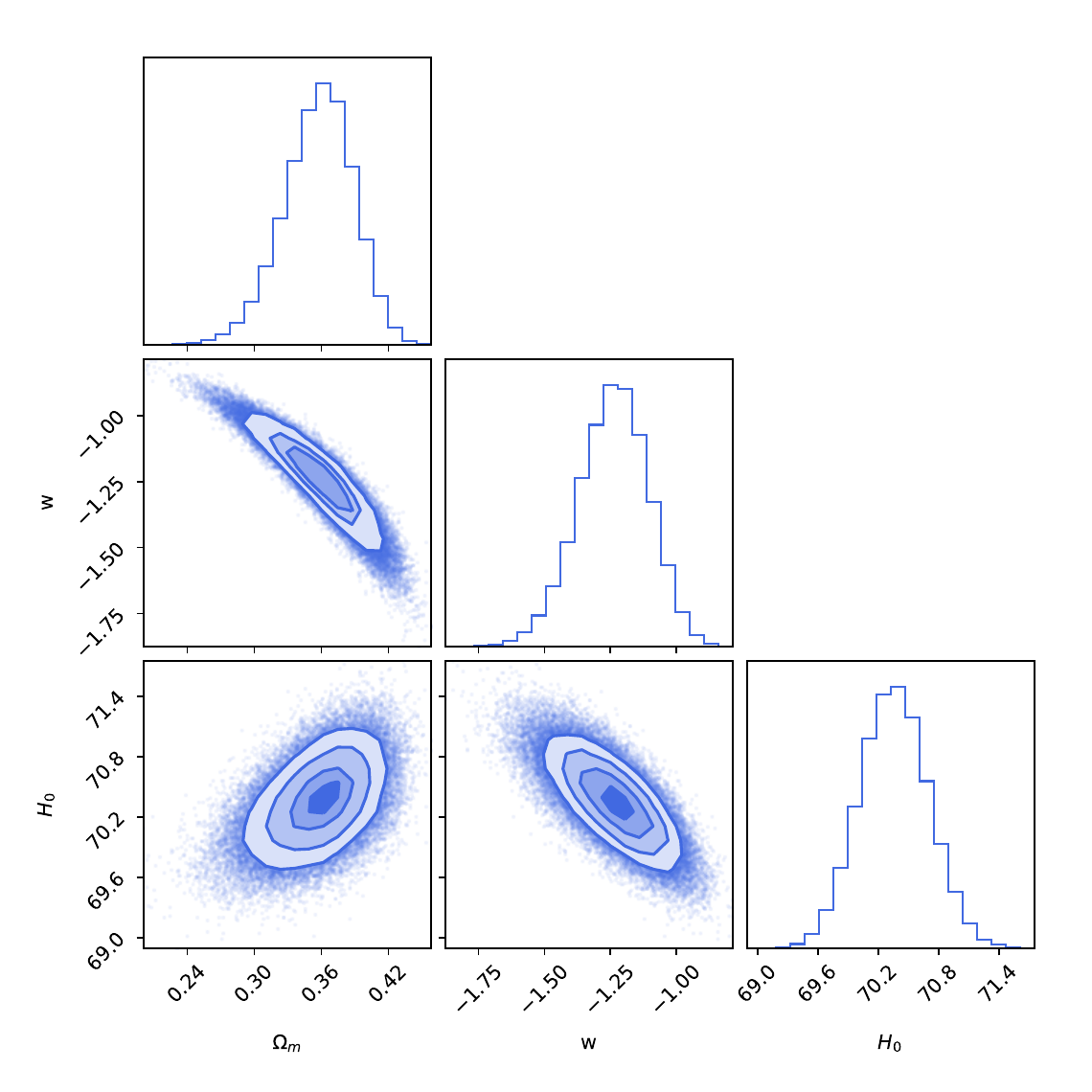}}\resizebox{80mm}{!}{\includegraphics[]{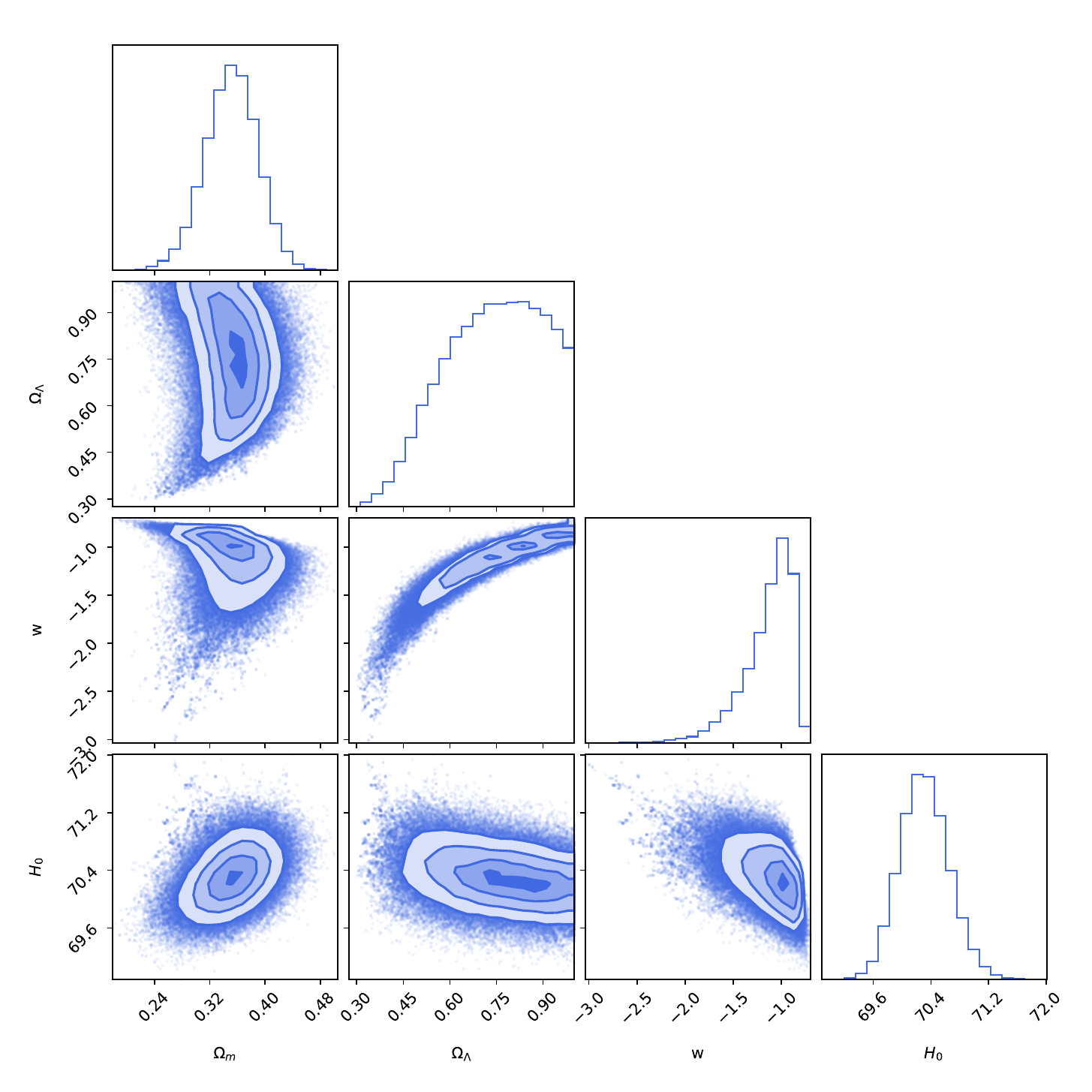}}\\
\caption{Constraints on the parameters of $\omega$CDM cosmological model using the SNe Ia data and the $L_{b,z}-T_{b,z}$ correlation of X-ray and optical samples of GRBs.
The left (right) is plotted for the flat (non-flat) $\omega$CDM model.}
\label{fig:L-T_wCDM_Ia+X+O_limit}
\end{figure*}

\begin{figure*}[ht!]\
\center
\resizebox{80mm}{!}{\includegraphics[]{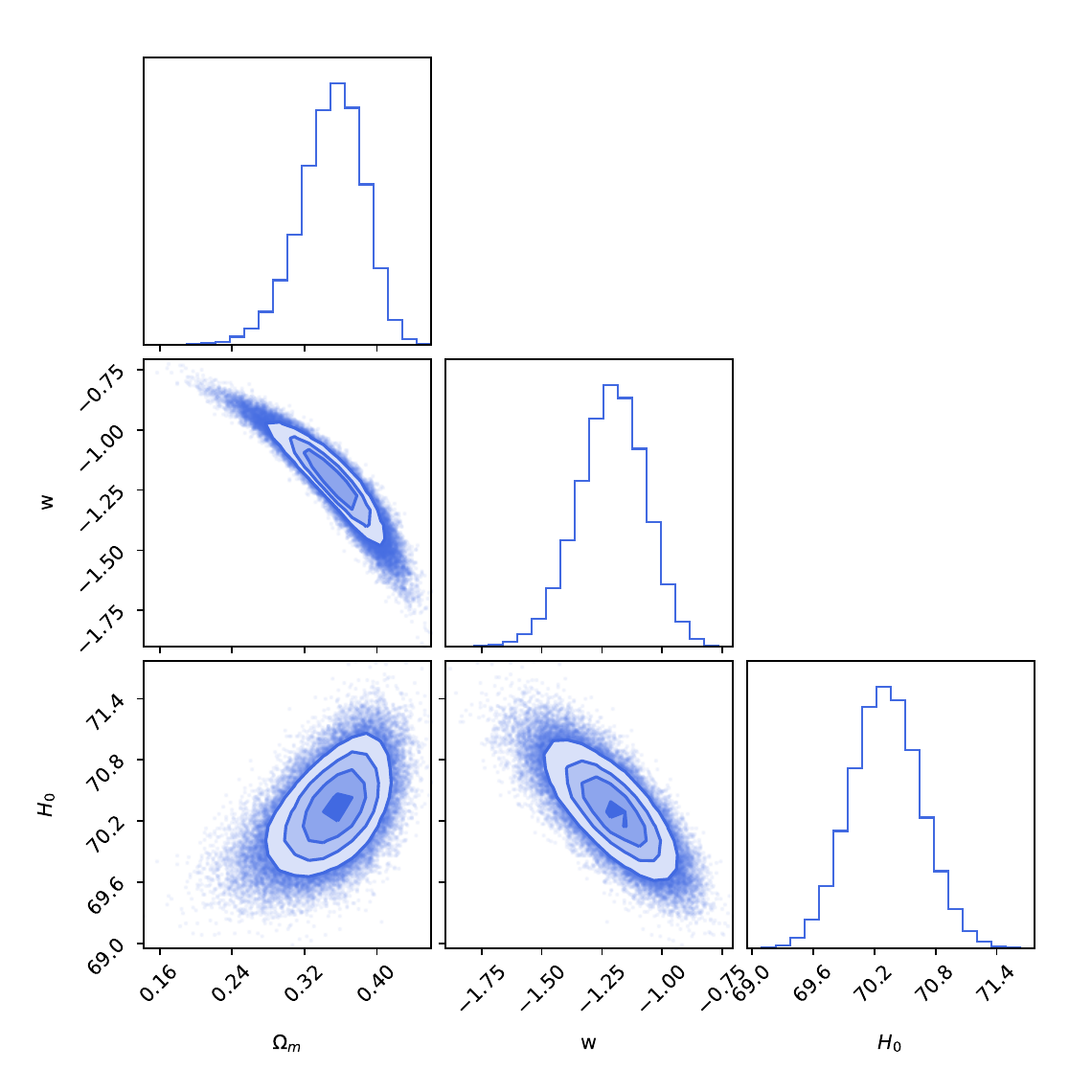}}\resizebox{80mm}{!}{\includegraphics[]{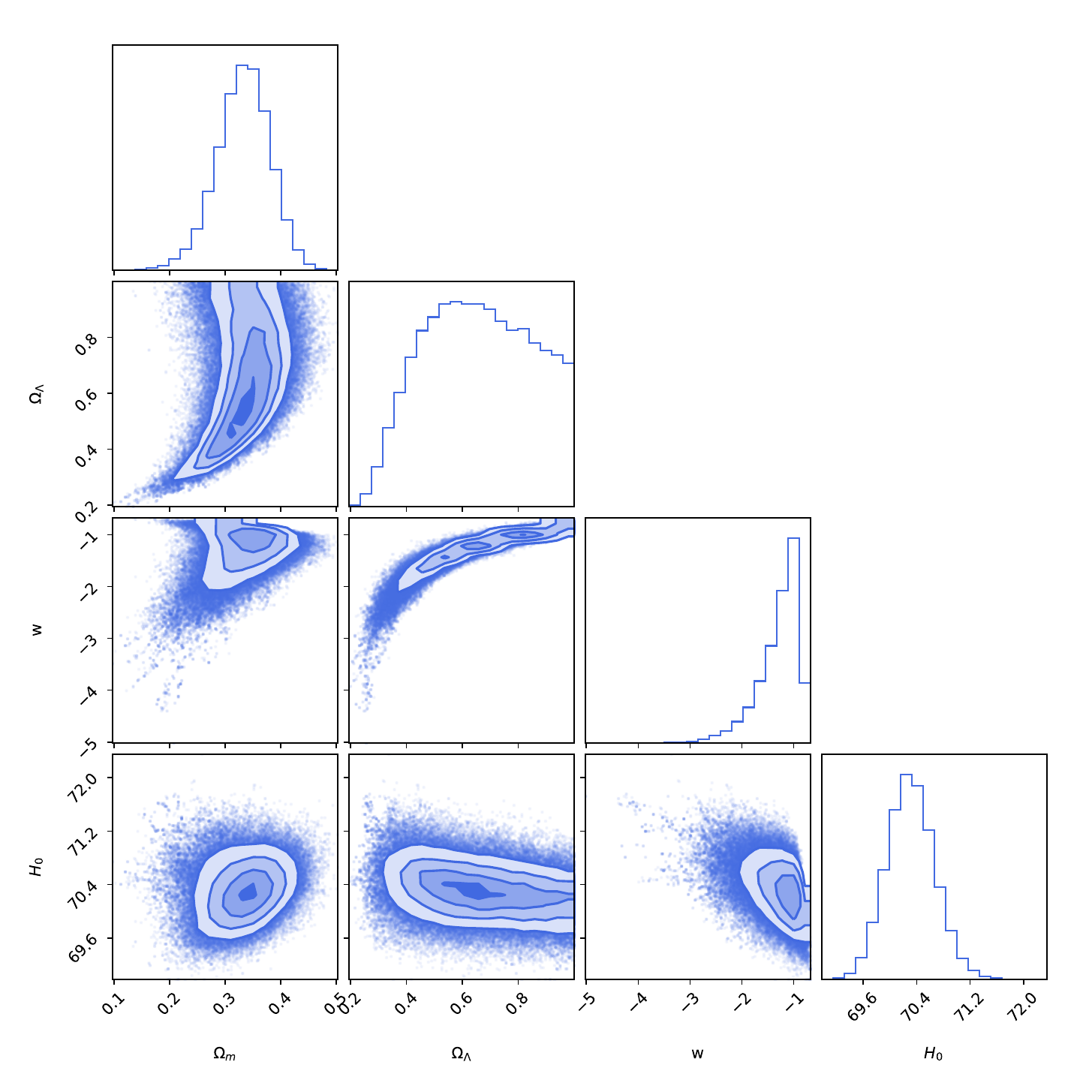}}\\
\caption{Constraints on the parameters of $\omega$CDM cosmological model using the SNe Ia data and the $L_{b,z}-T_{b,z}-E_{p,i}$ correlation of X-ray and optical samples of GRBs. The left (right) is plotted for the flat (non-flat) $\omega$CDM model.}
\label{fig:L-T-E_wCDM_Ia+X+O_limit}
\end{figure*}

Using the $L_{b,z}-T_{b,z}$ correlation, we obtain the constraints on the parameters. For a flat $\omega$CDM model with a single component of DE, combining the data of SNe Ia, X-ray and optical samples, the final results are $\Omega_{m} = 0.359_{-0.033}^{+0.029}$, $\omega = -1.235_{-0.136}^{+0.127}$ and $H_0 = 70.36_{-0.337}^{+0.343}$. For non-flat $\omega$CDM model, the constraints on the related parameters are $\Omega_{m} = 0.351_{-0.039}^{+0.036}$, $\Omega_{\Lambda} = 0.756_{-0.188}^{+0.162}$, $\omega = -1.072_{-0.326}^{+0.171}$ and $H_0 = 70.30_{-0.326}^{+0.345}$. The 1D and 2D distribution of the related parameters are shown in Fig. \ref{fig:L-T_wCDM_Ia+X+O_limit}. We also constrained the $\omega$CDM model by using the $L_{b,z}-T_{b,z}-E_{p,i}$ correlation. For flat model, the final constraints are $\Omega_{m} = 0.352_{-0.036}^{+0.032}$, $\omega = -1.206_{-0.143}^{+0.133}$ and $H_0 = 70.31_{-0.339}^{+0.345}$, and $\Omega_{m} = 0.331_{-0.054}^{+0.047}$, $\Omega_{\Lambda} = 0.615_{-0.210}^{+0.244}$, $\omega = -1.265_{-0.621}^{+0.311}$ and $H_0 = 70.35_{-0.353}^{+0.388}$ for non-flat model. The 1D and 2D distribution of the related parameters are shown in Fig. \ref{fig:L-T-E_wCDM_Ia+X+O_limit}. All of the results are summarized in Table \ref{tab:cos_para}. From the results, it is noted that the EOS parameter is close to the standard value $\omega = -1$, but the uncertainty is slightly large. In addition, the constraints on the related parameters are consistent with that of \cite{2022MNRAS.516.2575J}.

\begin{figure*}[ht!]\
\center
\resizebox{80mm}{!}{\includegraphics[]{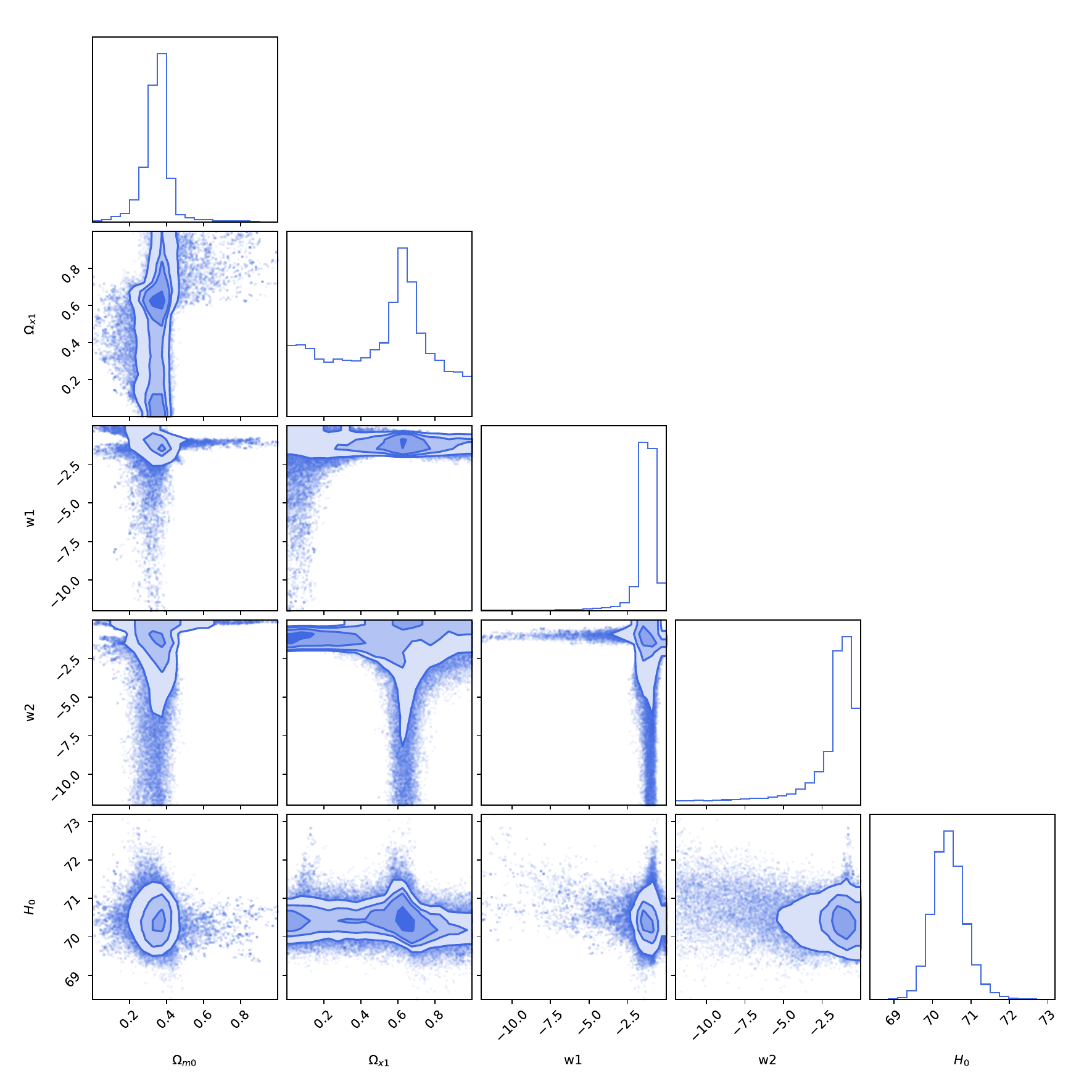}}\resizebox{80mm}{!}{\includegraphics[]{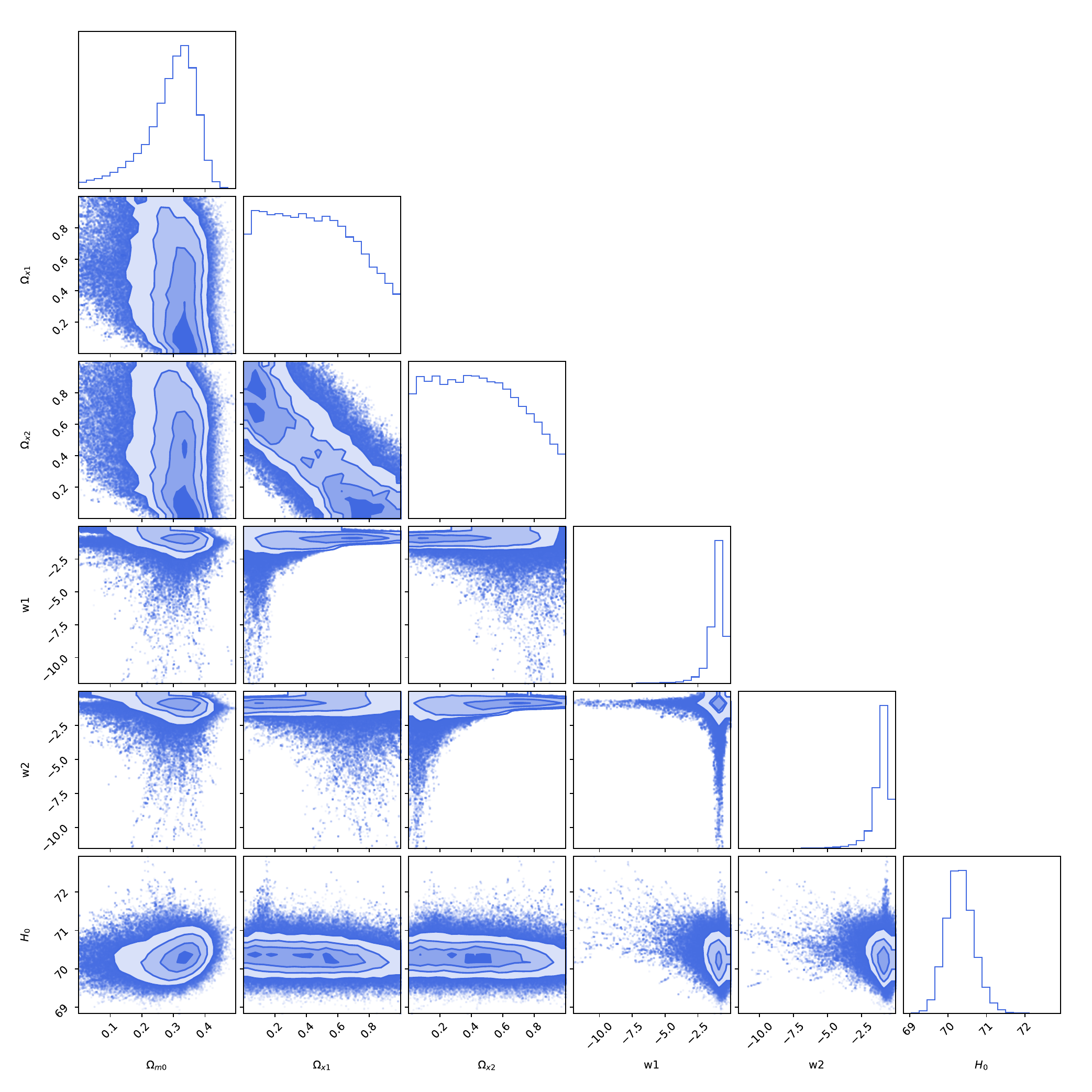}}\\
\caption{Constraints on the cosmological parameters on $X_1X_2$CDM model using the SNe Ia data and the $L_{b,z}-T_{b,z}$ correlation of X-ray and optical samples of GRBs.
The left (right) is plotted for the flat (non-flat) $X_1X_2$CDM model.}
\label{fig:L-T_X1X2CDM_Ia+X+O_limit}
\end{figure*}

\begin{figure*}[ht!]\
\center
\resizebox{80mm}{!}{\includegraphics[]{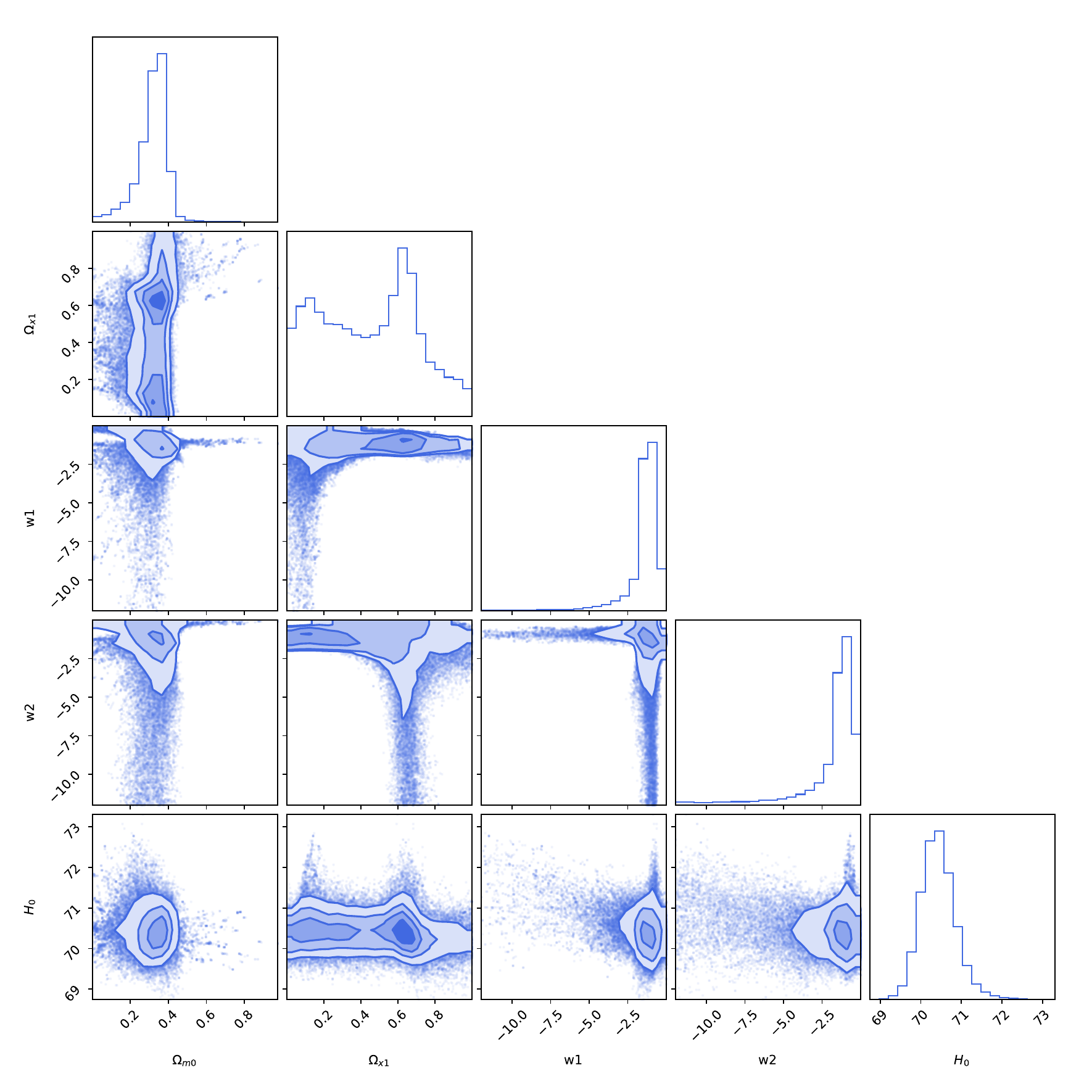}}\resizebox{80mm}{!}{\includegraphics[]{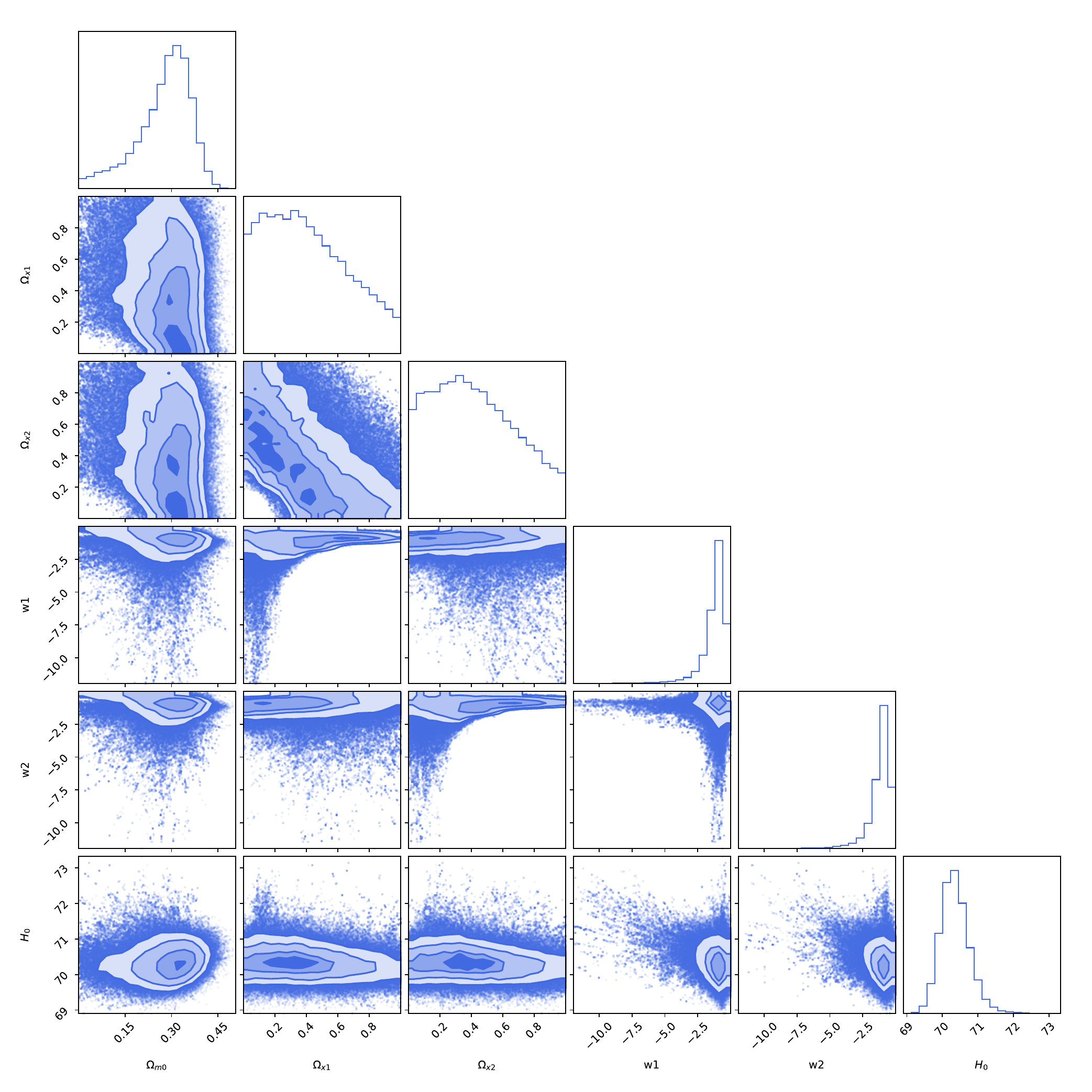}}\\
\caption{Constraints on the cosmological parameters in $X_1X_2$CDM model using the SNe Ia data and the $L_{b,z}-T_{b,z}-E_{p,i}$ correlation of X-ray and optical samples of GRBs.
The left (right) is plotted for the flat (non-flat) $X_1X_2$CDM model.}
\label{fig:L-T-E_X1X2CDM_Ia+X+O_limit}
\end{figure*}

\begin{figure*}[ht!]\
\center
\resizebox{90mm}{!}{\includegraphics[]{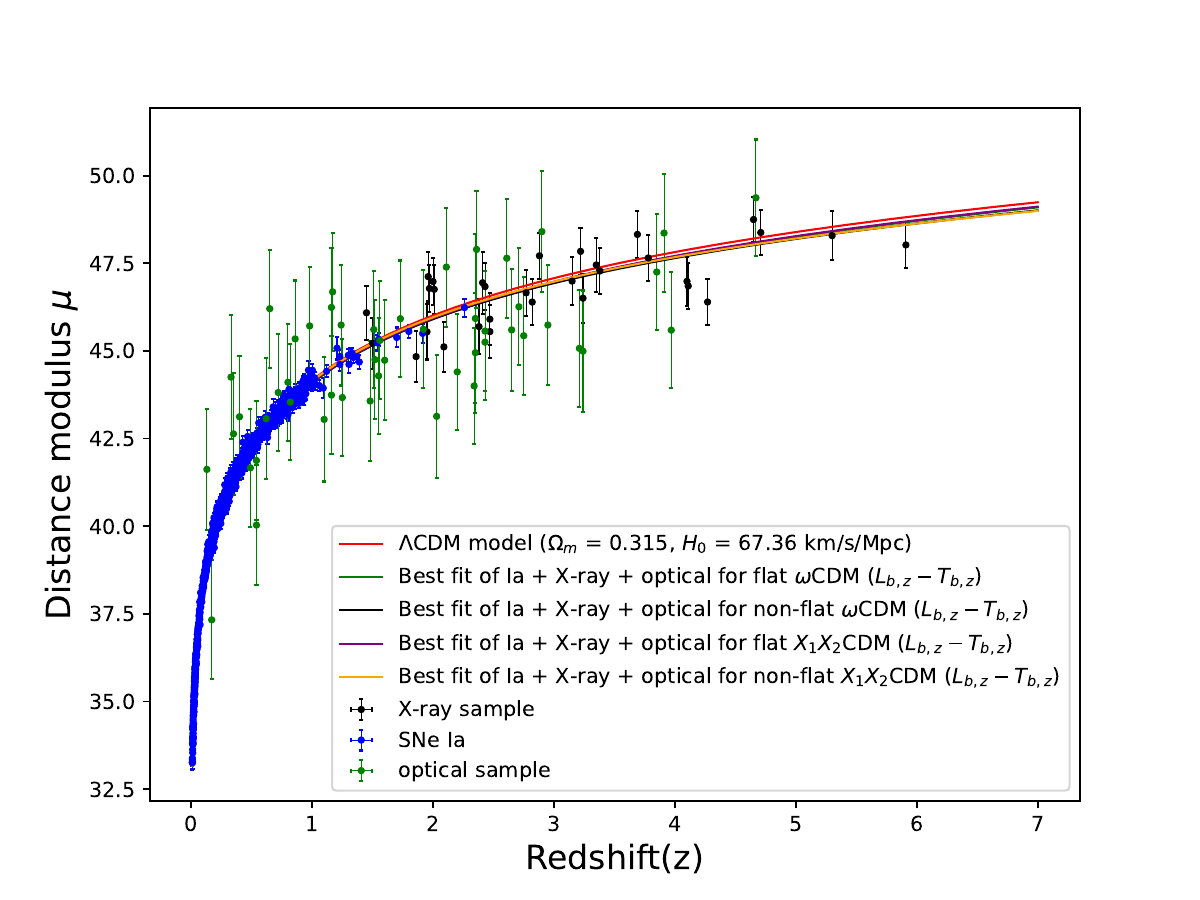}}\resizebox{90mm}{!}{\includegraphics[]{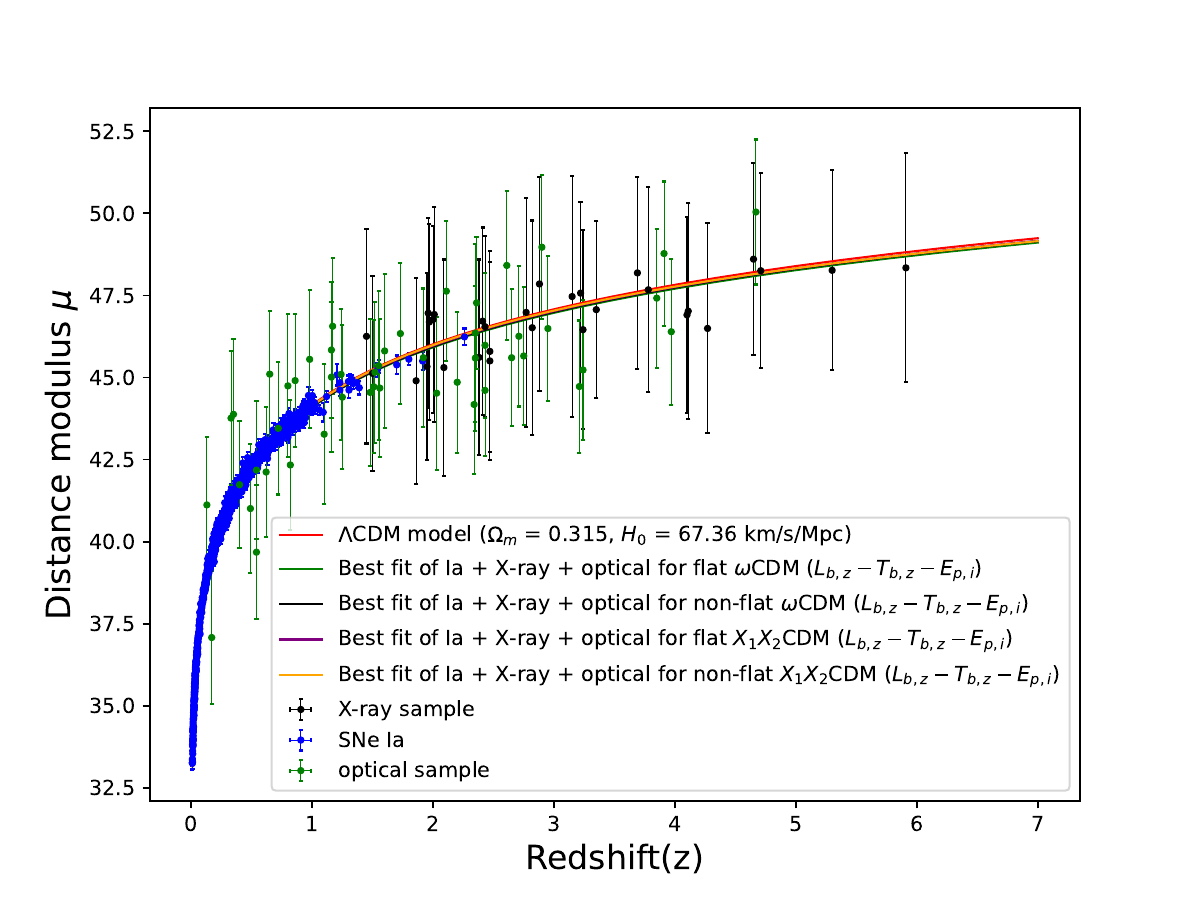}}
\caption{Calibrated GRB Hubble diagram at high redshift using the $L_{b,z}-T_{b,z}$ correlation (left-hand panel) and the $L_{b,z}-T_{b,z}-E_{p,i}$ correlation (right-hand panel) with SNe Ia, X-ray and optical samples. The black and green points represent X-ray and optical samples, respectively. Blue points are SNe Ia from the Pantheon sample.
The solid red line corresponds to the theoretical distance modulus calculated for a flat $\Lambda$CDM model
with $H_0 = 67.36~\rm km~s^{-1}~Mpc^{-1}$ and $\Omega_m = 0.315$.}
\label{fig:Hubble diagram}
\end{figure*}
%%%%%%%%%%%%%%%%%%%%%%%%%%%%%%%%%%%%%%%%%%%%%%%%%%%%%%%%%%%%%%%%%%%%%%%%%%%%

\begin{table*}[htbp]
  \begin{center}
  \caption{The best fitting results of the parameters for different correlations}
  %  \begin{threeparttable}

    \begin{tabular*}{0.86\linewidth}{c|c|c|c|c}
  \hline
  \hline
    $L_{b,z}-T_{b,z}$ correlation   & a & b &  $\sigma_{\rm int}$ \\
  \hline
  X-ray sample & $51.810 \pm 0.321$ & $-1.026 \pm 0.088$ &  $0.304 \pm 0.042$\\
  Optical sample & $48.849 \pm 0.346$ & $-0.979 \pm 0.094$ & $0.625 \pm 0.065$ \\
  \hline
   Calibrated $L_{b,z}-T_{b,z}$ correlation \\
  \hline
  X-ray sample & $51.696 \pm 0.540$& $-1.013 \pm 0.138$& $0.366 \pm 0.085$\\
  Optical sample & $49.045 \pm 0.419$& $-1.051 \pm 0.112$& $0.650 \pm 0.080$\\
  \hline
  $L_{b,z}-T_{b,z} - E_{p,i}$ correlation   & $a$ & $b$ & $c$ & $\sigma_{\rm int}$ \\
  \hline
  X-ray sample & $51.031 \pm 0.547$& $-0.974 \pm 0.091$&$0.230 \pm 0.134$ & $0.297 \pm 0.044$\\
  Optical sample & $47.453 \pm 0.592$& $-0.900 \pm 0.091$& $0.480 \pm 0.168$ & $0.573 \pm 0.061$\\
  \hline
   Calibrated $L_{b,z}-T_{b,z} - E_{p,i}$ correlation  \\
  \hline
  X-ray sample &$51.185 \pm 1.021$& $-1.000 \pm 0.143$& $0.188 \pm 0.325$ & $0.378 \pm 0.093$\\
  Optical sample &$47.565 \pm 0.623$& $-0.979 \pm 0.105$& $0.539 \pm 0.177$ & $0.574 \pm 0.074$\\
  \hline
  \hline
    \end{tabular*}
  \label{tab:Fitting}%
\end{center}
 % \end{threeparttable}
\end{table*}%

\begin{table*}[htbp]
  \caption{Constraints on the cosmological parameters of different models for different GRB samples and correlations. The AIC and BIC values of different cases are also shown.}
  \centering % centering table
  \resizebox{\textwidth}{60mm}{
    \begin{tabular}{c|c|c|c|c|c|c|c|c}
  \hline
  \hline
    Flat $\omega$CDM model   & $\Omega_m$  & $\omega$ & $H_0$ & $BIC$ & $AIC$\\
  \hline
  SNe Ia+X-ray ($L_{b,z}-T_{b,z}$)& $0.353_{-0.034}^{+0.030}$ & $-1.211_{-0.137}^{+0.126}$ & $70.32_{-0.333}^{+0.340}$ & 1120.00 & 1105.05\\
  SNe Ia+optical ($L_{b,z}-T_{b,z}$) & $0.356_{-0.036}^{+0.031}$ & $-1.222_{-0.145}^{+0.133}$ & $70.34_{-0.338}^{+0.349}$ & 1125.33 & 1110.31\\
  SNe Ia+X-ray+optical ($L_{b,z}-T_{b,z}$) & $0.359_{-0.033}^{+0.029}$ & $-1.235_{-0.136}^{+0.127}$ & $70.36_{-0.337}^{+0.343}$ & 1159.20 & 1144.11\\
  SNe Ia+X-ray ($L_{b,z}-T_{b,z}-E_{p,i}$)& $0.350_{-0.038}^{+0.033}$ & $-1.197_{-0.145}^{+0.135}$ & $70.30_{-0.339}^{+0.348}$ & 1087.70 & 1072.75\\
  SNe Ia+optical ($L_{b,z}-T_{b,z}-E_{p,i}$)& $0.352_{-0.038}^{+0.032}$ & $-1.205_{-0.144}^{+0.136}$ & $70.31_{-0.338}^{+0.345}$ & 1108.36 & 1093.36\\
  SNe Ia+X-ray+optical ($L_{b,z}-T_{b,z}-E_{p,i}$) & $0.352_{-0.036}^{+0.032}$ & $-1.206_{-0.143}^{+0.133}$ & $70.31_{-0.339}^{+0.345}$ & 1109.97 & 1094.88\\
  \hline
  Non-flat $\omega$CDM model  & $\Omega_m$  & $\Omega_{\Lambda}$ & $\omega$ & $H_0$ & $BIC$ & $AIC$\\
  \hline
  SNe Ia+X-ray ($L_{b,z}-T_{b,z}$) & $0.343_{-0.039}^{+0.037}$ & $0.718_{-0.189}^{+0.181}$ & $-1.109_{-0.371}^{+0.201}$ & $70.29_{-0.335}^{+0.361}$ & 1127.53 & 1107.60\\
  SNe Ia+optical ($L_{b,z}-T_{b,z}$) & $0.343_{-0.051}^{+0.044}$ & $0.673_{-0.223}^{+0.221}$ & $-1.182_{-0.534}^{+0.248}$ & $70.35_{-0.342}^{+0.376}$ & 1133.71 & 1113.71\\
  SNe Ia+X-ray + optical ($L_{b,z}-T_{b,z}$) & $0.351_{-0.039}^{+0.036}$ & $0.756_{-0.188}^{+0.162}$ & $-1.072_{-0.326}^{+0.171}$ & $70.30_{-0.326}^{+0.345}$ & 1165.99 & 1145.88\\
  SNe Ia+X-ray ($L_{b,z}-T_{b,z}-E_{p,i}$) & $0.321_{-0.060}^{+0.049}$ & $0.571_{-0.201}^{+0.262}$ & $-1.337_{-0.733}^{+0.365}$ & $70.37_{-0.366}^{+0.408}$ & 1096.22 & 1076.29\\
  SNe Ia+optical ($L_{b,z}-T_{b,z}-E_{p,i}$) & $0.330_{-0.057}^{+0.047}$ & $0.611_{-0.214}^{+0.245}$ & $-1.267_{-0.650}^{+0.310}$ & $70.36_{-0.356}^{+0.391}$ & 1116.78 & 1096.77\\
  SNe Ia+X-ray+optical ($L_{b,z}-T_{b,z}-E_{p,i}$) & $0.331_{-0.054}^{+0.047}$ & $0.615_{-0.210}^{+0.244}$ & $-1.265_{-0.621}^{+0.311}$ & $70.35_{-0.353}^{+0.388}$& 1118.90 & 1098.79\\
  \hline
  Flat $X_1X_2$CDM model   & $\Omega_{m0}$  &  $\Omega_{x1}$  &  $\omega_1$ & $\omega_2$ & $H_0$ & $BIC$ & $AIC$\\
  \hline
  SNe Ia+X-ray ($L_{b,z}-T_{b,z}$) & $0.338_{-0.078}^{+0.049}$ & $0.535_{-0.400}^{+0.214}$ & $-1.205_{-0.535}^{+0.412}$ & $-1.240_{-1.470}^{+0.728}$ & $70.37_{-0.366}^{+0.378}$ & 1137.91 & 1112.99\\
  SNe Ia+optical ($L_{b,z}-T_{b,z}$) & $0.336_{-0.100}^{+0.053}$ & $0.515_{-0.387}^{+0.225}$ & $-1.233_{-0.682}^{+0.490}$ & $-1.236_{-1.461}^{+0.684}$ & $70.39_{-0.366}^{+0.390}$ & 1149.66 & 1124.65\\
  SNe Ia+X-ray + optical ($L_{b,z}-T_{b,z}$) & $0.348_{-0.076}^{+0.050}$ & $0.553_{-0.423}^{+0.218}$ & $-1.237_{-0.524}^{+0.368}$ & $-1.211_{-1.514}^{+0.802}$ & $70.40_{-0.369}^{+0.379}$ & 1175.04 & 1149.90\\
  SNe Ia+X-ray ($L_{b,z}-T_{b,z}-E_{p,i}$) & $0.324_{-0.108}^{+0.058}$ & $0.515_{-0.380}^{+0.217}$ & $-1.195_{-0.703}^{+0.503}$ & $-1.205_{-1.531}^{+0.715}$ & $70.37_{-0.381}^{+0.399}$ & 1113.08 & 1088.16\\
  SNe Ia+optical ($L_{b,z}-T_{b,z}-E_{p,i}$) & $0.328_{-0.104}^{+0.056}$ & $0.516_{-0.387}^{+0.218}$ & $-1.207_{-0.666}^{+0.493}$ & $-1.213_{-1.549}^{+0.713}$ & $70.39_{-0.377}^{+0.391}$ & 1132.63 & 1107.62\\
  SNe Ia+X-ray+optical ($L_{b,z}-T_{b,z}-E_{p,i}$) & $0.326_{-0.108}^{+0.057}$ & $0.533_{-0.399}^{+0.205}$ & $-1.192_{-0.647}^{+0.517}$ & $-1.251_{-1.622}^{+0.735}$ & $70.37_{-0.369}^{+0.387}$ & 1135.70 & 1110.56\\
  \hline
  non-flat $X_1X_2$CDM model   & $\Omega_{m0}$  &  $\Omega_{x1}$  & $\Omega_{x2}$ & $\omega_1$ & $\omega_2$ & $H_0$ & $BIC$ & $AIC$\\
  \hline
  SNe Ia+X-ray ($L_{b,z}-T_{b,z}$) & $0.294_{-0.114}^{+0.059}$ & $0.396_{-0.288}^{+0.338}$ & $0.403_{-0.301}^{+0.343}$ & $-1.030_{-0.907}^{+0.497}$ & $-1.002_{-0.802}^{+0.493}$ & $70.35_{-0.359}^{+0.397}$ & 1157.23 & 1127.33\\
  SNe Ia+optical ($L_{b,z}-T_{b,z}$) & $0.288_{-0.114}^{+0.068}$ & $0.386_{-0.274}^{+0.351}$ & $0.412_{-0.292}^{+0.362}$ & $-1.015_{-1.010}^{+0.503}$ & $-0.983_{-0.911}^{+0.461}$ & $70.37_{-0.354}^{+0.404}$ & 1155.65 & 1125.64\\
  SNe Ia+X-ray+optical ($L_{b,z}-T_{b,z}$) & $0.295_{-0.121}^{+0.063}$ & $0.436_{-0.323}^{+0.339}$ & $0.414_{-0.310}^{+0.341}$ & $-0.959_{-0.758}^{+0.473}$ & $-0.994_{-0.782}^{+0.483}$ & $70.33_{-0.339}^{+0.386}$ & 1198.98 & 1168.81\\
  SNe Ia+X-ray ($L_{b,z}-T_{b,z}-E_{p,i}$)& $0.278_{-0.113}^{+0.068}$ & $0.335_{-0.255}^{+0.348}$ & $0.360_{-0.255}^{+0.348}$ & $-1.135_{-1.155}^{+0.578}$ & $-1.053_{-1.084}^{+0.572}$ & $70.38_{-0.362}^{+0.418}$ & 1119.34 & 1089.45\\
  SNe Ia+optical ($L_{b,z}-T_{b,z}-E_{p,i}$)& $0.279_{-0.115}^{+0.067}$ & $0.325_{-0.31}^{+0.337}$ & $0.365_{-0.254}^{+0.354}$ & $-1.133_{-1.181}^{+0.615}$ & $-1.049_{-1.073}^{+0.552}$ & $70.39_{-0.367}^{+0.413}$ & 1137.60 & 1107.59\\
  SNe Ia+X-ray+optical ($L_{b,z}-T_{b,z}-E_{p,i}$)& $0.284_{-0.111}^{+0.066}$ & $0.337_{-0.242}^{+0.358}$ & $0.376_{-0.266}^{+0.340}$ & $-1.113_{-1.178}^{+0.572}$ & $-1.047_{-0.992}^{+0.510}$ & $70.38_{-0.360}^{+0.409}$ & 1141.46 & 1111.29\\
  \hline
  \hline
    \end{tabular}}%
  \label{tab:cos_para}%

\end{table*}

We also constrained the $X_1X_2$CDM model by using the $L_{b,z}-T_{b,z}$ correlation and the final results can be found in Fig. \ref{fig:L-T_X1X2CDM_Ia+X+O_limit} and Table \ref{tab:cos_para}. For flat universe, the constraints are $\Omega_{m0} = 0.348_{-0.076}^{+0.050}$, $\Omega_{x1} = 0.553_{-0.423}^{+0.218}$, $\omega_1 = -1.237_{-0.524}^{+0.368}$, $\omega_2 = -1.211_{-1.514}^{+0.802}$ and $H_0 = 70.40_{-0.369}^{+0.379}$. For non-flat $X_1X_2$CDM model, the limits are $\Omega_{m0} = 0.295_{-0.121}^{+0.063}$, $\Omega_{x1} = 0.436_{-0.323}^{+0.339}$, $\Omega_{x2} = 0.414_{-0.310}^{+0.341}$, $\omega_1 = -0.959_{-0.758}^{+0.473}$, $\omega_2 = -0.994_{-0.782}^{+0.483}$ and $H_0 = 70.33_{-0.339}^{+0.386}$. By using the $L_{b,z}-T_{b,z}-E_{p,i}$ correlation, with the SNe Ia, X-ray and optical samples, we obtained $\Omega_{m0} = 0.326_{-0.108}^{+0.057}$, $\Omega_{x1} = 0.533_{-0.399}^{+0.205}$, $\omega_1 = -1.192_{-0.647}^{+0.517}$, $\omega_2 = -1.251_{-1.622}^{+0.735}$ and $H_0 = 70.37_{-0.369}^{+0.387}$ for the flat model. For non-flat model, we obtained $\Omega_{m0} = 0.284_{-0.111}^{+0.066}$, $\Omega_{x1} = 0.337_{-0.242}^{+0.358}$, $\Omega_{x2} = 0.376_{-0.266}^{+0.340}$, $\omega_1 = -1.113_{-1.178}^{+0.572}$, $\omega_2 = -1.047_{-0.992}^{+0.510}$ and $H_0 = 70.38_{-0.360}^{+0.409}$, which can be seen in Fig. \ref{fig:L-T-E_X1X2CDM_Ia+X+O_limit} and Table \ref{tab:cos_para}. By analyzing the results of the derived data, we find that the error of the two-component model of the equation of state parameters $\omega_1$ and $\omega_2$ are larger than that of the parameters $\omega$ in the one-component model, especially the lower limit error. We can see that the distribution has a long tail when $\omega$ is smaller than -1 from the one-dimensional distribution of $\omega_1$ and $\omega_2$ in the Figs. \ref{fig:L-T_X1X2CDM_Ia+X+O_limit} and \ref{fig:L-T-E_X1X2CDM_Ia+X+O_limit}, and the peak distributions of $\omega_{1,2}$ are around -1 for both of flat and non-flat cases. These results are consistent with that shown in Fig. 4 of \cite{2007PhRvD..76l3007G}.

The matter density $\Omega_{m0}$ is almost the same for the one-component and two-component DE models. The one-dimensional histograms of $\Omega_{x1}$ in the Figs. \ref{fig:L-T_X1X2CDM_Ia+X+O_limit} and \ref{fig:L-T-E_X1X2CDM_Ia+X+O_limit} show an additional relatively small peak for the flat two-component DE model. For this model, one have $\Omega_{m0}+ \Omega_{x1} + \Omega_{x2} = 1$, and our results show that $\Omega_{m0}$ is close to 0.3 and the maximum probability of $\Omega_{x1}$ is around 0.6. Then the possible density of $\Omega_{x2}$ is close to 0.1, which should be the optimal solution of the likelihood function. Given the two-component DE model, the optimal solution is likewise satisfied when $\Omega_{x2}$ is close to 0.6 and $\Omega_{x1}$ is close to 0.1. Therefore, there should be a relatively large number of data points at 0.1, resulting in the two peaks in the histogram. These results are consistent with those reported in Figs. 3, 5, and 7 by \cite{2007PhRvD..76l3007G}.

We also combine X-ray and optical samples with SNe Ia data, respectively, to constrain the $\omega$CDM and $X_1X_2$CDM models using the two and three-parameter correlations. All of the results are summarized in Table \ref{tab:cos_para}. It can be found that the constraints on the related parameters are consistent for the same GRB correlation. For the same cosmological model, $\Omega_m$ and $\Omega_x$ constrained using the three-parameter correlation are slightly smaller than that derived from the two-parameter correlation. Although there is a very significant error bar, we note that the peaks in the $\omega_1$ and $\omega_2$ distributions show a similar feature.

In Fig. \ref{fig:Hubble diagram}, we present Hubble diagram of different GRBs samples calibrated using the $L_{b,z}-T_{b,z}$ correlation (left) and the $L_{b,z}-T_{b,z}-E_{p,i}$ correlation (right). It is worth noting that the error bars for the distance modulus derived from the two correlations for high redshift GRBs are still large. These results suggest that the calibrated correlations of GRBs alone cannot currently measure the cosmological parameters as precisely as SNe Ia. However, GRBSs are valuable for expanding the Hubble diagram to high redshift. It is expected that GRBs can be employed as a full-fledged cosmic probe rather than an auxiliary tool when the frequency of high-redshift events seen increases and forthcoming observations precision improves.

We also used AIC and BIC to compare different models, they can be written as

\begin{equation}\label{eq:AIC}
{\rm AIC} = \chi^2_{min} + 2p,
\end{equation}
and
\begin{equation}\label{eq:BIC}
{\rm BIC} = \chi^2_{min} + p{\rm ln}(N),
\end{equation}
where $\chi^2$ is defined in Eq. \ref{eq:chi^2}. $N$ is the number of data points and $p$ is the number of free parameters of the relative model. The model with lower AIC or BIC values is favored by the criterion, and we compute $\rm \Delta X$ (where $\rm X = \rm AIC~or~\rm BIC$) values with respect to the flat $\omega$CDM model. A negative $\rm \Delta X$ suggests that the model under consideration performs better than the reference model. Normally, if it is positive, one can refer to: $\rm \Delta X \in [0,2]$ indicates the candidate model has substantial support for the reference model, and $\rm \Delta X \in (2,6]$ indicates the candidate model receives less support than the reference model. The candidate model is not strongly supported for $\rm \Delta X > 6$. The values of AIC and BIC for $\omega$CDM and $X_1X_2$CDM models are shown in Table \ref{tab:cos_para}.  Comparing the results of the models by the information criteria, we find that AIC and BIC always favor models with less free parameters.

%%%%%%%%%%%%%%%%%%%%%%%%%%%%%%%%%%%%%%%%%%%%%%%%%%%%%%%%%%%%%%%%%%%%%%%%%

\section{Discussion and conclusion}
\label{section:5}

~~~~Based on the features of the light curves, we have screened GRBs with a plateau phase followed by a decay phase with a decay index of -2 . The optical samples with a platform and decay phase are screened using the same procedure. It has been suggested that GRBs screened with this characteristic should have same physical origin. Therefore, we have acquired both X-ray and optical samples for our studies. We found that the selected optical samples have a compact $L_{b,z}-T_{b,z}$ correlation as the X-ray samples selected by \cite{2022ApJ...924...97W}. We have investigated the $L_{b,z}-T_{b,z}-E_{p,i}$ correlation for the two sets of samples, and found that the internal dispersion ($\sigma_{\rm int}$) of the three-parameter correlation is smaller than that of the two-parameter correlation for the same sets of samples, indicating that the $L_{b,z}-T_{b,z}-E_{p,i}$ correlation is tighter than the $L_{b,z}-T_{b,z}$ correlation.

For the purpose of cosmological probe with GRBs, we have applied the GP method to calibrate the two correlation and obtained the model-independent distance modulus. Then the MCMC approach was utilized to derive the best fitting results of the calibrated correlations. The $\sigma_{\rm int}$ of the calibrated correlations were found to be slightly larger, which could be due to less GRB samples at lower redshifts. It is expected that the GRB database will be enlarged in near future due to more observers entering service, allowing us to do more extensive screening and studies.

Using the calibrated X-ray and optical samples, we have constrained the free parameters of the one-component DE model ($\omega$CDM model) and two-component DE model ($X_1X_2$CDM model). We found that the selected optical and X-ray sample can effectively constrain the cosmological parameters, showing that careful screening of GRBs is necessary for cosmological purposes.  The restriction results of the three-parameter correlation for both groups of samples accord with the two-parameter correlation. The constraints on $\omega_1$ and $\omega_2$ of the $X_1X_2$CDM model are also consistent at $1\sigma$ level. It seems to imply that the composition of DE prefers a single component. We also used the AIC and BIC information criteria to compare the different models, and found that the model with fewer free parameters is favoured.

Previous studies, for example, \cite{2007PhRvD..76l3007G}, \cite{2018ApJ...863...50S}, \cite{2022ApJ...924...97W}, \cite{2023ApJ...953...58L}, have investigated the correlation of X-ray samples with a platform phase, as well as the correlation of optical samples, cosmological applications, and two-component DE model. Compared with previous works, our studies have the following new characteristics: (1) We have selected different samples of GRBs that could have the same physical mechanism based on the decay index, and updated the constrains on the cosmological parameters of the $X_1X_2$CDM model using more SNe Ia and GRBs data. (2) We have investigated not only the $L_{b,z}-T_{b,z}$ connection existed in the selected GRBs data, but also the much tighter $L_{b,z}-T_{b,z}-E_{p,i}$ correlation, and have applied both of the correlations for the cosmological probes.

\section*{Acknowledgments}
This work is supported by the National Natural Science Foundation of China (Grant Nos. U2038106 and 12273009), the Shandong Provincial Natural Science Foundation (Grant No. ZR2021MA021), Jiangsu Funding Program for Excellent Postdoctoral Talent (20220ZB59) and Project funded by China Postdoctoral Science Foundation (2022M721561).

\end{document}